\documentclass[twoside]{article}
\usepackage{qic,epsfig,bbm,mathtools}

\newtheorem{thm}{Theorem}
\newtheorem{de}[thm]{Definition}
\newtheorem{lem}[thm]{Lemma}
\newtheorem{prop}[thm]{Propositon}

\begin{document}
\setlength{\textheight}{8.0truein}    

\runninghead{Golden Codes}
            {V. Londe, A. Leverrier}

\normalsize\textlineskip
\thispagestyle{empty}
\setcounter{page}{1}

\copyrightheading{0}{0}{2003}{000--000}

\vspace*{0.88truein}

\alphfootnote

\fpage{1}

\centerline{\bf
GOLDEN CODES: QUANTUM LDPC CODES}
\vspace*{0.035truein}
\centerline{\bf BUILT FROM REGULAR TESSELLATIONS OF HYPERBOLIC 4-MANIFOLDS}
\vspace*{0.37truein}
\centerline{\footnotesize
VIVIEN LONDE}
\vspace*{0.015truein}
\centerline{\footnotesize\it Inria Paris, 2 rue Simone Iff}
\baselineskip=10pt
\centerline{\footnotesize\it Paris, 75012, France}
\centerline{\footnotesize\it vivien.londe@gmail.com}
\vspace*{10pt}
\centerline{\footnotesize 
ANTHONY LEVERRIER}
\vspace*{0.015truein}
\centerline{\footnotesize\it Inria Paris, 2 rue Simone Iff}
\baselineskip=10pt
\centerline{\footnotesize\it Paris, 75012, France}
\centerline{\footnotesize\it anthony.leverrier@inria.fr}
\publisher{January, $8^{th}$ 2018}{March, $8^{th}$ 2019}

\vspace*{0.21truein}

\abstracts{
We adapt a construction of Guth and Lubotzky \cite{GL14} to obtain a family of quantum LDPC codes with non-vanishing rate and minimum distance scaling like $n^{0.1}$ where $n$ is the number of physical qubits. Similarly as in Ref.~\cite{GL14}, our homological code family stems from hyperbolic 4-manifolds equipped with tessellations. The main novelty of this work is that we consider a regular tessellation consisting of hypercubes. We exploit this strong local structure to design and analyze an efficient decoding algorithm. 
}{}{}

\vspace*{10pt}

\keywords{quantum information, quantum error correction, homological quantum codes, hyperbolic geometry, arithmetic manifolds.}
\vspace*{3pt}
\communicate{to be filled by the Editorial}

\vspace*{1pt}\textlineskip    

\section{Introduction}        

Building a large-scale quantum computer is certainly one of the main challenges faced by the physics community in the 21th century. This turns out to be a daunting task because of the extreme fragility of quantum information: any uncontrolled interaction between the qubits and the environment leads to decoherence and quickly causes any computation to fail. 
Fortunately, theoretical solutions exist under the form of quantum error correcting codes which allow one to encode logical qubits into a larger number of physical qubits, in such a way that logical information can be preserved and recovered despite potential errors occurring on the physical qubits \cite{Sho95,CS96}.\\

Mathematically, a quantum code of dimension $k$ and length $n$ is a subspace of $(\mathbbm{C}^2)^{\otimes n}$ of dimension $2^k$. A possible way to specify such a subspace is \textit{via} a stabilizer group: an Abelian subgroup of the $n$-qubit Pauli group. In that case, the quantum code is defined as the common eigenspace of the stabilizers with eigenvalue $+1$. Such a code is called a \textit{stabilizer code} \cite{Got97}.
Among these, CSS codes, due to Calderbank, Shor and Steane, are those for which the stabilizer group admits a list of $n-k$ generators which are either products of $X$-type Pauli operators, or of $Z$-type Pauli operators. A convenient way to find CSS codes, initiated by Kitaev \cite{KIT03}, is to consider tessellations of manifolds. 
In that case, the physical qubits are associated with $i$-dimensional faces of the tessellation, the $X$-type stabilizers are associated with $(i-1)$-faces and the $Z$-type stabilizers with $(i+1)$-faces. 
Such codes are called \emph{homological}. The stabilizer group of a homological code is commutative as required because ($i+1$)-faces, $i$-faces and ($i-1$)-faces form a chain complex. In other words, given an $(i+1)$-face $\mathcal{P}$, and an $(i-1)$-face $\mathcal{Q}$, there is an even number of $i$-faces incident to both $\mathcal{P}$ and $\mathcal{Q}$ \cite{BM07, Zem09, BT16}.\\

A major advantage of homological codes with a fixed and compact local structure is that they are naturally of the low-density parity-check (LDPC) type, meaning that generators of the stabilizer group act nontrivially on a constant number of qubits and that each qubit is acted upon by a constant number of generators. This is of course especially interesting for potential implementations, but also at a more mathematical level since classical LDPC codes play a central role in classical coding theory. 
A second advantage of homological codes is that they can lead to simple and efficient decoding algorithms which directly exploit the local structure of the code on the manifold \cite{DKLP02, Har04, DZ17, DN17}.\\

The parameters $[[n,k,d]]$ of homological codes can be derived from the properties of the underlying manifold: the length $n$ of the code is given by the number of $i$-faces in the tessellation, the dimension $k$ is given by the rank of the $i^{\mathrm{th}}$ homology group, and the minimum distance, that is the minimum weight of a nontrivial Pauli error, is related to the $i^{\mathrm{th}}$ homological systole of the manifold, that is the minimal number of $i$-faces forming a homologically nontrivial $i$-cycle.   
Exploiting this connection with manifolds exhibiting systolic freedom, Freedman, Meyer and Luo \cite{FML02} were able to construct the quantum LDPC codes with the best minimum distance presently known, achieving $d = \Theta( n^{1/2} \log^{1/4} n)$ \footnote{Ref.~\cite{FML02} mentions $d = \Theta( n^{1/2} \log^{1/2} n)$ but it seems that there is a typo in their numerical application.}.\\

An important question is to understand what parameters $[[n,k, d]]$ can be achieved with quantum LDPC codes. The toric code and the code of Ref.~\cite{FML02} display a large minimum distance but only encode a constant number of qubits, $k= O(1)$. If the manifold is Euclidean, strong constraints apply on the code parameters: namely the parameters have to satisfy $k d^2 \leq c n$ for some constant $c$ \cite{BPT10}. For tessellations of 2-dimensional manifolds, Delfosse showed that $k d^2  \leq c (\log k)^2 n$ \cite{Del13}. In particular, these results show that one cannot get a good minimum distance for codes with constant rate built from 2-dimensional manifolds. \\

In many cases, it is natural to consider constant-rate codes where $k = \Theta(n)$: such codes for instance allow one to obtain quantum fault-tolerant computation with constant space overhead \cite{Got13, FGL18}. For a long time, it was believed that constant-rate homological codes could not have a large minimum distance, that is growing polynomially with their length \cite{Zem09}. A recent breakthrough was the work of Guth and Lubotzky \cite{GL14} who gave a construction of homological codes in hyperbolic 4-space that combine a constant rate with a minimum distance $d = n^{\alpha}$, for some constant $\alpha >0$. It was later shown by Murillo that the construction could be adapted to yield $\alpha \in [0.2, 0.3]$.
Quickly after this result, Hastings proposed a decoding algorithm for such codes \cite{Has16}. Unfortunately, the analysis of Hastings' decoding algorithm is only valid when local brute-force decoding is performed at a scale that may not be computationally practical.
In fact, it is difficult to precisely analyze the performance of Hastings' decoder because the local structure of the codes of \cite{GL14} is not completely explicit.\\

In this work, we give a variant of the construction of Guth and Lubotzky which admits a simple explicit local structure: it is based on a regular tessellation of the 4-dimensional hyperbolic space by means of hypercubes. We then exploit this local structure to design an efficient decoding algorithm which tries to locally shorten cycles. 
In Section \ref{sec:variant}, we give an overview of our approach compared to that of Guth and Lubotzky. 
In Section \ref{sec:cubes}, we explain how to obtain a regular tessellation of hyperbolic $4$-space with hypercubes. 
In Section \ref{sec:compact}, we detail how to quotient the space to get a compact manifold, which then yields the quantum code. 
We finally describe our local decoder and analyze its performances in Section \ref{sec:decoding}.

\section{A variant of Guth and Lubotzky's construction based on a Regular Tessellation of Hyperbolic Space}
\label{sec:variant}

The family of manifolds considered in \cite{GL14} is a family of 4-dimensional hyperbolic coverings. The tessellations can be obtained by pulling back the natural tessellation of the base space. Each covering equipped with its natural tessellation gives rise to a quantum error correcting code. Unfortunately the fundamental polytope of this natural tessellation is not regular. In particular, it is nontrivial to obtain the local structure of the tessellation, and therefore an explicit description of the code generators. While this did not prevent Hastings from designing a decoding algorithm for these codes \cite{Has16}, simulating its performance for the codes of \cite{GL14} appears quite impractical. (Note, however, that Hastings' decoder was recently implemented for the 4-dimensional toric code, in Euclidean space \cite{BDMT16}.)\\

It is useful to see the 4-dimensional homological quantum error correcting codes that Guth and Lubotzky and we construct as generalisations of the 2-dimensional toric code. Let us therefore give the arithmetic manifold viewpoint on the toric code. 
We consider the ordinary tessellation of the Euclidean plane by unit squares such that vertices have integer coordinates. The translation group of Euclidean plane is $\mathbbm{R} \times \mathbbm{R}$. We denote by $\Gamma_{toric}$ the subgroup $\mathbbm{Z} \times \mathbbm{Z}$ of this translation group. Elements of $\Gamma_{toric}$ stabilize the ordinary tessellation of Euclidean plane. Let $I = p\mathbbm{Z}$ be an ideal of $\mathbbm{Z}$, with $p$ a positive integer and define $\Gamma(I)_{toric}$ to be $I \times I$. The quotient $\mathcal{M}_{toric}(I)$ of the Euclidean plane by $\Gamma(I)_{toric}$ is a torus, that naturally inherits the tessellation by unit squares from the Euclidean plane. 
The constructions of \cite{GL14} and of the present work are generalisations of the 2-dimensional Euclidean toric code in a 4-dimensional hyperbolic setting. To help the reader draw analogies with the toric code, we introduced in this paragraph notations similar to the notations used in the sequel. \\

We now summarise the construction of Guth and Lubotzky and explain the advantages of our approach. In Ref.~\cite{GL14}, the construction is based on tessellations of the hyperbolic 4-space $\mathbbm{H}^4$. To each code corresponds a manifold equipped with a tessellation. The base space $\mathcal{M}$ is constructed by considering the action of a cocompact discrete group of isometries $\Gamma$ on hyperbolic 4-space: $\mathcal{M} = \Gamma \backslash \mathbbm{H}^4$. To each finite index subgroup $\Gamma(I)$ of $\Gamma$ corresponds a covering $\mathcal{M}(I)$ of $\mathcal{M}$ given by $\mathcal{M}(I) = \Gamma(I) \backslash \mathbbm{H}^4$. It is natural to tessellate $\mathcal{M}$ with a single 4-face and to tessellate $\mathcal{M}(I)$ with a number of 4-faces equal to the index of $\Gamma(I)$ in $\Gamma$. Each 4-face is isometric to the first one. Unfortunately the 4-face is not regular in \cite{GL14}, which makes the local description of the quantum code rather complicated.  To obtain a similar construction with a regular 4-face, we reverse the process: we start with a convenient regular 4-face and then build a corresponding discrete group of isometries $\Gamma$. \\

For its symmetries and because it tessellates the hyperbolic 4-space, we choose the 4-dimensional hypercube as our targeted regular 4-face. We embed it in hyperbolic 4-space and scale it according to the $\{4,3,3,5\}$ tessellation of hyperbolic 4-space (see Section \ref{subsubsec:schlafli} for a definition of Sch\"afli symbols). The group $\Gamma$ is generated by the direct isometries of hyperbolic 4-space sending opposing faces of the hypercube to each other with no rotation. The tessellating 4-face we obtain is a hypercube by construction. \\

The tricky part of the construction is to define finite index subgroups of our discrete group of isometries $\Gamma$ in a way similar to \cite{GL14}. Indeed, arithmeticity of subgroups $\Gamma(I)$ plays a central role in lower bounding the minimum distance of the corresponding error correcting codes. To achieve this goal, we rely on arithmetic structures defined over the number field  $\mathbbm{Q}(\sqrt{5})$. Replacing $\mathbbm{Q}$ by this number field, $\mathbbm{Z}$ by $\mathbbm{Z}[\phi]$ - the ring of integers of $\mathbbm{Q}(\sqrt{5})$ - and ideals $p\mathbbm{Z}$ by ideals of $\mathbbm{Z}[\phi]$ where $\phi$ is the \emph{golden ratio} (giving its name to our construction), it is possible to define principal congruence subgroups $\Gamma(I)$ such that the corresponding family of error correcting codes satisfies the same asymptotic estimates as in \cite{GL14}. We therefore obtain a family of codes with a regular local structure, a non-vanishing rate and a minimum distance lower bounded by $n^{0.1}$, where $n$ is the number of physical qubits. \\

We take advantage of the regular local structure to design an efficient decoding algorithm. This algorithm is highly local. For X errors, it decreases the syndrome at the scale of a single 4-face. In particular, our algorithm is more local and explicit than Hastings' decoder \cite{Has16}. We prove that syndromes associated with errors of weight below the injectivity radius of the manifold always contain a pattern that can be locally shortened so as to decrease the weight of the syndrome. In other words, the algorithm simply consists in examining the syndrome in the neighborhood of an error and acting on qubits to decrease the syndrome weight. We show that arbitrary errors of size $O(\log n)$ are corrected by this algorithm, which in turn implies that random errors will be corrected with high probability if the error rate is small enough. These results are similar to those of Hastings' decoder, but with the advantage of an entirely explicit algorithm with precise bounds on its performances. 

\section{Hyperbolic 4-space and its Regular Tessellation by Hypercubes} 
\label{sec:cubes}

In this section, we first introduce the minimal background on hyperbolic 4-space and regular tessellations. We then focus on the tessellation of hyperbolic 4-space by 4-dimensional hypercubes on which our quantum code construction is based.

\subsection{Hyperbolic space}

We use the hyperboloid model to describe 4-dimensional hyperbolic space. As a set, 4-dimensional hyperbolic space is identified with 
\[\mathbbm{H}^{4} = \{(x_{0},x_{1},x_{2},x_{3},x_{4}) \in \mathbbm{R}^{5} / \\ -x_{0}^{2}+x_{1}^{2}+x_{2}^{2}+x_{3}^{2}+x_{4}^{2} = -1, \quad x_{0} > 0 \}.\]
It is endowed with a Riemannian metric such as to make it a space of constant negative sectional curvature. Its orientation-preserving isometry group is $SO^{o}(1,4)$, the identity component of the special indefinite orthogonal group. The four coordinates $x_{1}$, $x_{2}$ ,$x_{3}$ and $x_{4}$ are sufficient to parametrise $\mathbbm{H}^{4}$. Indeed $x_{0}$ can be retrieved from the condition $x_{0}^{2} = 1+x_{1}^{2}+x_{2}^{2}+x_{3}^{2}+x_{4}^{2}$. Therefore in the sequel we will ignore the coordinate $x_{0}$ and refer to $x_{1}$ as the \textit{first} coordinate and not as the second. \\

The reader is referred to \cite{Rat06} for a comprehensive introduction to hyperbolic geometry. To give some intuition about hyperbolic space we will merely compare the perimeter growth of a hyperbolic circle of radius $r$ with its Euclidean counterpart. In hyperbolic space, such a circle has perimeter $2\pi \sinh(r)$. The growth is exponentially faster than its Euclidean counterpart $2\pi r$. In spherical space on the other hand, the perimeter of a circle of radius $r$ is only $2\pi \sin(r)$, for $r < \pi$. Informally speaking, there is more room in the angular direction in hyperbolic space than in Euclidean space just like there is less room in the angular direction in spherical space than in Euclidean space. One can make this statement more precise by considering regular tessellations and their combinatorial properties.

\subsection{Regular polytopes and tessellations}

The geometric point of view on tessellations is probably the most intuitive. By geometric, we mean that vertices, edges and higher dimensional faces of the tessellation are subsets of a geometric space such as for example the hyperbolic plane or the Euclidean 3-space. However a tessellation also entails purely combinatorial data, namely the incidences between its i-faces and its (i+1)-faces. We will refer to this combinatorial data as the abstract polytope attached to a tessellation. For a comprehensive exposition of this so-called abstract point of view we refer to the book of McMullen and Schulte \cite{MS02} (Chapter 2 for the abstract point of view and Chapter 5 for its interplay with geometric realisations). The abstract point of view is especially relevant to quantum error correction since the combinatorial data is sufficient to define a quantum error correcting code. We will only mention here that an abstract polytope is called regular if its automorphism group is transitive on the set of its flags. Moreover the realisation of a abstract regular polytope as a tessellation of a geometric space is called regular if its automorphism group can be represented as an isometry group of the geometric space.  \\

Interestingly the combinatorial data of a abstract regular polytope (its incidences) determines in which geometric space it can be embedded. We can thus talk about spherical, Euclidean and hyperbolic abstract regular polytopes: \\

\begin{de} \label{def1}
An abstract regular polytope is called \emph{spherical} (respectively \emph{Euclidean}, respectively \emph{hyperbolic}) if it can be realised with regular faces in a spherical (respectively Euclidean, respectively hyperbolic) manifold. \\
\end{de}

Informally speaking, if a polytope is locally too small to fit in Euclidean space, it curves inwards and yields a spherical tessellation. If it is too big, it yields a hyperbolic tessellation.\\
In the Euclidean case, the faces of the tessellation can be scaled by multiplying all lengths by a given positive real $\lambda$. In the spherical and hyperbolic cases, however, the volumes of faces are imposed by the combinatorics of the tessellation: tessellations far from being Euclidean lead to faces with a large volume.\\

\subsubsection{Combinatorial point of view on tessellations: Schl{\"a}fli symbols} 
\label{subsubsec:schlafli}

Results of this section come from Ref.~\cite{Cox54}. Since the realisation of an abstract regular polytope is essentially unique (up to a scaling factor if it is euclidean) we will often not distinguish a regular tessellation of a geometric space from its abstract regular polytope: the combinatorial data attached to it. Therefore we can describe regular tessellations \textit{via} their \emph{Schl{\"a}fli symbols}, which are defined recursively for $p, q, r, \dots$ positive integers:
\begin{itemize}
\item \{$p$\} refers to a regular $p$-sided polygon.
\item \{$p$,$q$\} refers to a regular tessellation by regular $p$-sided polygons such that each vertex is incident to $q$ regular $p$-sided polygons. \\
One obtains a tessellation of the Euclidean plane if $(p-2)(q-2) = 4$, or of the hyperbolic plane if $(p-2)(q-2) > 4$. Finally if $(p-2)(q-2) < 4$, then $\{p,q\}$ can represent either a tessellation of the two-dimensional sphere or a 3-dimensional polyhedron. \\
There are five regular 3-dimensional polyhedrons called the \emph{Platonic solids}: the regular \emph{icosahedron} $\{3,5\}$;  the regular \emph{octahedron} $\{3,4\}$; the regular \emph{tetrahedron} $\{3,3\}$; the \emph{cube} $\{4,3\}$ and the regular \emph{dodecahedron}, $\{5,3\}$. 

\item If \{$p$,$q$\} and \{$r$,$q$\} are 3-dimensional polyhedrons\footnote{The condition that $\{r,q\}$ is also a 3-dimensional polyhedron is necessary, for instance, to ensure that the dual tessellation $\{r,q,p\}$ is well-defined.}, then \{$p$,$q$,$r$\} refers to a regular tessellation by \{$p$,$q$\}-polyhedrons such that each edge of the tessellation is incident to $r$ \{$p$,$q$\}-polyhedrons. Note that the terminology \emph{honeycomb} is sometimes used instead of tessellation to insist on 3-dimensionality. The terminology \emph{mosaic} can be encountered as well. We will use tessellation in the sequel regardless of the dimension. \\
Similarly as before, the nature of the tessellation depends on the relation between the integers $p, q, r$. 
If $\cos\! \big(\frac{\pi}{q}\big) = \sin\!\big(\frac{\pi}{p}\big)\sin\! \big(\frac{\pi}{r}\big)$, one obtains a tessellation of the Euclidean 3-dimensional space.
If $\cos\! \big(\frac{\pi}{q}\big) > \sin\! \big(\frac{\pi}{p}\big)\sin\! \big(\frac{\pi}{r}\big)$, one gets a tessellation of the hyperbolic 3-dimensional space.
Finally, if $\cos\! \big(\frac{\pi}{q}\big) < \sin\!\big(\frac{\pi}{p}\big)\sin\! \big(\frac{\pi}{r}\big)$, it can represent either a tessellation of the spherical 3-dimensional space or a 4-dimensional polytope. \\
There are six regular 4-dimensional polytopes: \{3,3,5\} is the \emph{600-cell}, \{3,3,4\} is the \emph{4-orthoplex}, \{3,4,3\} is the \emph{24-cell}, \{3,3,3\} is the regular \emph{4-simplex}, \{4,3,3\} is the 4-dimensional \emph{hypercube}, and \{5,3,3\} is the \emph{120-cell}.

 \item If \{$p$,$q$,$r$\} and \{$s$,$r$,$q$\} are 4-dimensional polytopes, \{$p$,$q$,$r$,$s$\} refers to a regular tessellation by \{$p$,$q$,$r$\}-polytopes such that each 2-face of the tessellation is incident to $s$ \{$p$,$q$,$r$\}-polytopes. \\
 If $\frac{\cos^{2}(\frac{\pi}{q})}{\sin^{2}(\frac{\pi}{p})} + \frac{\cos^{2}(\frac{\pi}{r})}{\sin^{2}(\frac{\pi}{s})} = 1$, 
 it is a tessellation of the 4-dimensional Euclidean space. 
 If $\frac{\cos^{2}(\frac{\pi}{q})}{\sin^{2}(\frac{\pi}{p})} + \frac{\cos^{2}(\frac{\pi}{r})}{\sin^{2}(\frac{\pi}{s})} > 1$, it is a tessellation of hyperbolic 4-space. \\
There are five regular tessellations of hyperbolic 4-space: \{3,3,3,5\}, \{4,3,3,5\}, \{5,3,3,5\}, \{5,3,3,4\} and \{5,3,3,3\}. \\
\end{itemize}

Given a tessellation or a polytope described by Schl{\"a}fli symbol \{$p_{1}$,...,$p_{n}$\}, the tessellation or polytope described by \{$p_{n}$,...,$p_{1}$\} is called the \emph{dual} tessellation or polytope. It is the tessellation obtained by mapping every $i$-face to an $(n-i)$-face. Note that duality doesn't change the hyperbolic, Euclidean or spherical type of a tessellation. 

\subsubsection{The \{4,3,3,5\} regular tessellation of hyperbolic 4-space }

In this work we will focus on the \{4,3,3,5\} regular tessellation of hyperbolic 4-space. The 4-faces \{4,3,3\} of this tessellation are 4-dimensional hypercubes, which are especially convenient. In particular, one can exploit the fact that the symmetries of the hypercube are compatible with its description in coordinates in the hyperboloid model to find a nice description of a discrete subgroup of $SO^{o}(1,4)$ corresponding to the \{4,3,3,5\} regular tessellation. 
The  other regular tessellations of hyperbolic 4-space could lead to similar constructions and would yield quantum codes with similar asymptotic properties. Since their symmetries are less compatible with coordinates in the hyperboloid model, they would require more work to make computations explicit and we will not consider them in this work. 

\subsection{Isometry group of the tessellation}

We consider a regular hypercube centered at the origin of the hyperboloid model and such that each of its eight 3-faces is orthogonal to a coordinate axis. We denote this hypercube by $T$ in the sequel  (a 4-dimensional hypercube is also called a tesseract). Since the hypercube $T$ is regular, there exist direct isometries of the hyperbolic 4-space $\mathbbm{H}^{4}$ sending any one of them onto the opposite one. These isometries can be thought of as the hyperbolic equivalent of Euclidean translations. Requiring that these direct isometries act trivially on three coordinates defines them uniquely. For example the direct isometries that send the 3-faces orthogonal to the first coordinate axis onto each other are given by the following matrices:

\vspace{-10pt}
\begin{equation}
\notag
g_{1} = \begin{pmatrix}
\cosh t & \sinh t & 0 & 0 & 0 \\
\sinh t & \cosh t  & 0 & 0 & 0  \\
0 & 0 & 1 & 0 & 0 \\
0 & 0 & 0 & 1 & 0 \\
0 & 0 & 0 & 0 & 1 \\
\end{pmatrix}, \quad
(g_{1})^{-1} = \begin{pmatrix}
\cosh t  & -\sinh t & 0 & 0 & 0 \\
-\sinh t & \cosh t & 0 & 0 & 0  \\
0 & 0 & 1 & 0 & 0 \\
0 & 0 & 0 & 1 & 0 \\
0 & 0 & 0 & 0 & 1 \\
\end{pmatrix}.
\end{equation}

\noindent
The pair of direct isometries sending the 3-faces orthogonal to the second coordinate axis onto each other is given by these two matrices:

\vspace{-10pt}
\begin{equation}
\notag
g_{2} = \begin{pmatrix}
\cosh t & 0 & \sinh t & 0 & 0 \\
0 & 1  & 0 & 0 & 0  \\
\sinh t & 0 & \cosh t & 0 & 0 \\
0 & 0 & 0 & 1 & 0 \\
0 & 0 & 0 & 0 & 1 \\
\end{pmatrix}, \quad
(g_{2})^{-1} = \begin{pmatrix}
\cosh t  & 0 & -\sinh t & 0 & 0 \\
0 & 1 & 0 & 0 & 0  \\
-\sinh t & 0 & \cosh t & 0 & 0 \\
0 & 0 & 0 & 1 & 0 \\
0 & 0 & 0 & 0 & 1 \\
\end{pmatrix}.
\end{equation}

\noindent
The two remaining pairs of direct isometries $g_{3}$, $(g_{3})^{-1}$, $g_{4}$ and $(g_{4})^{-1}$ are obtained from the above matrices by permuting two coordinates. Recall that the coordinate $x_{0}$ is redundant with the four others. Therefore the zeroth coordinate should not be permuted. \\

The angle between two adjacent 3-faces of the hypercube $T$ depends on the volume of $T$ or equivalently on the parameter $t$: as we will show, the greater $t$, the greater the volume of $T$ and the smaller the angle between two adjacent 3-faces. We will compute the value of $t$ such that this angle is $2\pi/5$. Indeed in the \{4,3,3,5\} regular tessellation of hyperbolic 4-space, five hypercubes meet along each 2-face, which means that the \emph{dihedral angle} between two 3-faces of the same hypercube must be $2\pi/5$. Note that the dihedral angle between 3-faces is sometimes called dichoral angle to insist on higher dimension. In the sequel we will use the terminology dihedral angle regardless of dimension. To compute dihedral angles in the hyperboloid model we need some definitions. \\

\begin{de}[Ratcliffe \cite{Rat06} $\S 3.1$]
The \emph{Lorentzian inner product} denoted $\circ$ is the bilinear map defined on $\mathbbm{R}^{5} \times \mathbbm{R}^{5}$ by:
$$u \circ v = -u_{0}v_{0} + u_{1}v_{1} + u_{2}v_{2} + u_{3}v_{3} + u_{4}v_{4}.$$
Two vectors $u$, $v$ are \emph{Lorentz orthogonal} if $u \circ v = 0$. \\
\end{de}

\begin{de}[Ratcliffe \cite{Rat06} $\S 3.1$]
The \emph{Lorentzian norm} of a vector $u$ is the complex number denoted $||u||$ satisfying $u \circ u = ||u||^{2}$ and such that $||u||$ is either positive imaginary, $0$ or positive.\\
Note that if $||u||$ is positive imaginary, $|||u|||$ denotes its modulus. \\
\end{de} 

\begin{de}[Ratcliffe \cite{Rat06} $\S 3.2$]
The \emph{space-like angle} $\eta$ between two space-like vectors $u$ and $v$ is defined by: $u \circ v = ||u|| \cdot ||v|| \cos(\eta)$ and $0 \leq \eta \leq \pi$. \\
\end{de}

\begin{de}[Ratcliffe \cite{Rat06} $\S 6.4$]
Let $S$ and $T$ be two adjacent sides of a convex polytope $P$. Let $u$, respectively $v$, be a vector that is Lorentz orthogonal to $S$, respectively $T$, and directed away from $P$. Let $\eta$ the space-like angle between $u$ and $v$. Then the dihedral angle $\theta(S,T)$ between $S$ and $T$ is defined by:
$$ \theta(S,T) = \pi - \eta(u,v).$$
\end{de}

\noindent
With this definition the dihedral angle is invariant under global Lorentz transformations. Indeed Lorentz orthogonality and space-like angles are Lorentz invariant. This property is necessary since Lorentz transformations are the isometries of the hyperbolic metric. \\

To fully justify the definition it remains to show that we obtain the expected dihedral angle when the two sides intersect at the origin $(1,0,0)^{T}$ of the hyperboloid. For simplicity we assume that $S$ and $T$ are two lines intersecting at the origin of the hyperbolic plane. In the hyperboloid model we can assume that $S = \{(\cosh(x),\sinh(x),0) \, | \, x \in \mathbbm{R}  \}$ and $T = \{(\cosh(x),\cos(\theta(S,T))\sinh(x),\sin(\theta(S,T))\sinh(x)) \, | \, x \in \mathbbm{R}  \}$. Then $u = (0,0,-1)$ is Lorentz orthogonal to $S$ and $v = (0,-\sin(\theta(S,T)),\cos(\theta(S,T)))$ is Lorentz orthogonal to $T$. Vectors $u$ and $v$ are directed away from $P$ ( $P$ is defined consistently with $\theta(S,T)$ ). A computation yields $u \circ v = -\cos(\theta(S,T))$. Since $||u|| = ||v|| = 1$ we obtain that $\cos(\eta(u,v)) = - \cos(\theta(S,T))$. Since by definition both $\theta(S,T)$ and $\eta(u,v)$ are in $[ 0, \pi ]$ , this gives $\theta(S,T) = \pi - \eta(u,v)$. \\

We can now come back to the hypercube $T$ centered at the origin of hyperbolic space. Let $C_{1}$, respectively $C_{2}$, be the 3-face of $T$ orthogonal in hyperbolic 4-space to the first, respectively second, axis and such that its first, respectively second, coordinate in the hyperboloid model is positive. Recall that $x_{0}$ is referred to as the zeroth coordinate. Thus the first coordinate is $x_{1}$ and the second is $x_{2}$. Points of $C_{1}$ have coordinates of the form $\lambda (\cosh (t/2),\sinh(t/2),a,b,c)^{T}$ for some $a, b, c \in \mathbbm{R}$ and a normalising constant $\lambda$.  Similarly points of $C_{2}$ have coordinates of the form $\lambda (\cosh(t/2),a,$ $\sinh(t/2),b,c)^{T}$.
It is straightforward to verify that $N_{1} = (\sinh(t/2), \cosh(t/2), 0, 0, 0)^{T}$ is Lorentz orthogonal to $C_{1}$ and $N_{2} = (\sinh(t/2),0,\cosh(t/2),0,0)^{T}$ is Lorentz orthogonal to $C_{2}$. We have: 

\vspace{-10pt}
\begin{align*}
\eta(N_{1},N_{2}) &= \text{arccos}( \frac{N_{1} \circ N_{2}}{||N_{1}||||N_{2}||} )= \text{arccos}(-\sinh^{2}(t/2)), \\
\theta(C_{1},C_{2}) &= \pi - \eta(N_{1},N_{2}), \\
\theta(C_{1},C_{2}) &= \pi - \text{arccos}(-\sinh^{2}(t/2)).
\end{align*} 

\noindent
As announced, the dihedral angle between two adjacent 3-faces of the hypercube $T$ decreases with parameter $t$, or equivalently when the volume of $T$ increases. \\

Since we want to build a \{4,3,3,5\} tessellation, five hypercubes have to be incident to each 2-face of the hypercube. This imposes $\theta(C_{1},C_{2}) =  2\pi/5 $ and leads to $t = 2\text{arsinh}(\sqrt{\cos(2\pi/5)})$. We eventually obtain 
$$\cosh(t) = \frac{1+\sqrt{5}}{2} \quad \text{and} \quad \sinh(t) = \sqrt{\frac{1+\sqrt{5}}{2}},$$
the golden ratio $\phi$ and its square root. \\

We denote by $\Gamma$ the discrete subgroup of $SO^{o}(1,4)$ generated by the four direct isometries $g_{1}$, $g_{2}$, $g_{3}$ and $g_{4}$ sending a 3-face of the hypercube onto the opposite 3-face. Note that there are eight such direct isometries but they are pairwise inverse of each other.

\subsection{Coxeter group approach }
\label{subsec:coxeter}

The problem with the group $\Gamma$ defined above is that its fundamental domain is not the hypercube. Indeed there are elements of $\Gamma$ that fix the hypercube globally but not pointwise. They correspond to symmetries of the hypercube. Therefore the fundamental domain of $\Gamma$ is only a fraction of the hypercube. \\

However we would like to work with $\{4,3,3,5\}$ tessellations. This way the tessellation is regular with a local structure that is easy to describe. To achieve this goal we will define a double extension of $\Gamma$ which is a representation of the $\{4,3,3,5\}$ Coxeter group. The standard technique of considering cosets of this Coxeter group will then yield the tessellation by hypercubes (see \cite{MS02}). We can define $\Gamma^{\{4,3,3,5\}} =  \langle r_0, r_1, r_2, r_3, r_4 \rangle$, with its five generators given by: \\

\begin{align*}
&r_{0} = \begin{pmatrix}
1 & 0 & 0 & 0 & 0\\
0 & -1 & 0 & 0 & 0\\
0 & 0 & 1 & 0 & 0\\
0 & 0 & 0 & 1 & 0\\
0 & 0 & 0 & 0 & 1\end{pmatrix}, \qquad
r_{1} = \begin{pmatrix}
1 & 0 & 0 & 0 & 0\\
0 & 0 & 1 & 0 & 0\\
0 & 1 & 0 & 0 & 0\\
0 & 0 & 0 & 1 & 0\\
0 & 0 & 0 & 0 & 1\end{pmatrix}, \qquad
r_{2} = \begin{pmatrix}
1 & 0 & 0 & 0 & 0\\
0 & 1 & 0 & 0 & 0\\
0 & 0 & 0 & 1 & 0\\
0 & 0 & 1 & 0 & 0\\
0 & 0 & 0 & 0 & 1\end{pmatrix}, \\
&r_{3} = \begin{pmatrix}
1 & 0 & 0 & 0 & 0\\
0 & 1 & 0 & 0 & 0\\
0 & 0 & 1 & 0 & 0\\
0 & 0 & 0 & 0 & 1\\
0 & 0 & 0 & 1 & 0\end{pmatrix}, \qquad
r_{4} = \begin{pmatrix}
\phi & 0 & 0 & 0 & -\sqrt{\phi} \\
0 & 1 & 0 & 0 & 0\\
0 & 0 & 1 & 0 & 0\\
0 & 0 & 0 & 1 & 0\\
\sqrt{\phi} & 0 & 0 & 0 & -\phi \end{pmatrix}.
\end{align*}

\medskip

It is straightforward to verify that these five generators satisfy the relations defining the $\{4,3,3,5\}$ string Coxeter group:
$$r_0^2 = r_1^2 = r_2^2 = r_3^2 = r_4^2 = (r_0 r_1)^4 = (r_1 r_2)^3 = (r_2 r_3)^3 = (r_3 r_4)^5 = \text{id}.$$ 

\medskip

To obtain a $\{4,3,3,5\}$ tessellation of $\mathbbm{H}^4$, we can follow \cite{MS02} and identify $S_i$-cosets in $\Gamma^{\{4,3,3,5\}}$ with i-faces of the $\{4,3,3,5\}$ tessellation. For $i \in \{0,\ldots, 4\}$, the group $S_i$ is the subgroup of $\Gamma^{\{4,3,3,5\}}$ generated by the four generators $(r_j)_{j \neq i}$ (for instance $S_1 \stackrel{def}{=} \langle r_0, r_2, r_3, r_4 \rangle$). By definition, an $i$-face $F_a$ and a $j$-face $F_b$ are incident if the corresponding cosets $g_a S_i$ and $g_b S_j$ have a non-empty intersection. \\

With these definitions we only have a combinatorial description of the $\{4,3,3,5\}$ tessellation. To obtain the geometrical version of the tessellation, observe that each element of  $\Gamma^{\{4,3,3,5\}}$ corresponds to a simplex of $\mathbbm{H}^4$: the identity of $\Gamma^{\{4,3,3,5\}}$ corresponds to a fundamental domain $\mathcal{S}$ of $\mathbbm{H}^4$ (which happens to be a simplex in this case) and any $g \in \Gamma^{\{4,3,3,5\}}$ corresponds to $g \mathcal{S}$. Now an $i$-face of the tessellation corresponds to a coset of $S_i$ in $\Gamma^{\{4,3,3,5\}}$. This coset can be considered as a set of elements of $\Gamma^{\{4,3,3,5\}}$ or, in other words, as a set of closed simplices of $\mathbbm{H}^4$. If $i=4$ the geometrical 4-face is defined as the union of these closed simplices. If $i \in \{0,\ldots,3\}$ the geometrical i-face is defined as the intersection of the 4-faces incident to this $i$-face. \\

This $\{4,3,3,5\}$ tessellation of hyperbolic 4-space has an infinite number of $i$-faces for every $i \in \{0,\ldots,4\}$. To build a code with a finite number of qubits, we need a tessellation with a finite number of 2-faces. We use in the sequel number theoretical tools to construct quotients of the hyperbolic 4-space equipped with a $\{4,3,3,5\}$ tessellation. 

\section{Compact Manifolds equipped with a $\{4,3,3,5\}$ Tessellation}
\label{sec:compact}

We want to define a quantum code by identifying physical qubits with 2-faces of a tessellation. To obtain a code with a finite number of physical qubits, we will consider tessellations of compact manifolds. We will therefore consider the $\{4,3,3,5\}$ tessellation of compact manifolds obtained as quotients of hyperbolic 4-space. 
These manifolds are called \emph{arithmetic} because they are quotients of $\mathbbm{H}^{4}$ by arithmetic subgroups of $\Gamma^{\{4,3,3,5\}}$. We first review the definitions of a number field and its ring of integers. We then use these tools to associate an arithmetic subgroup $\Gamma^{\{4,3,3,5\}}(I)$ to every ideal $I$ of the ring of integers $\mathbbm{Z}[\phi]$. 

\subsection{Number fields and rings of integers}

\bigskip
\begin{de}
A \emph{number field} $K$ is a finite degree field extension of the field of rational numbers $\mathbbm{Q}$. \\
\end{de}

\begin{de}
A complex number is an \emph{algebraic number} if it is a root of a non-zero polynomial over $\mathbbm{Q}$. \\
\end{de}

\begin{thm}[\textit{e.g.} Marcus, \cite{Mar77} Appendix 2]
Every number field has the form $\mathbbm{Q}(\alpha)$ for some algebraic number $\alpha \in \mathbbm{C}$. If $\alpha$ is a root of an irreducible polynomial over $\mathbbm{Q}$ having degree n, then
$$ \mathbbm{Q}(\alpha) = \{a_{0} + a_{1}\alpha + ... + a_{n-1} \alpha^{n-1} \, | \, \forall i \in \{0,\ldots ,n-1\} , a_{i} \in \mathbbm{Q} \}. $$
\end{thm}

\noindent
Since $\sqrt{5}$ is a root of $X^{2}-5$, which is irreducible over $\mathbbm{Q}$, we have
$$\mathbbm{Q}(\sqrt{5}) = \{ a_{0} + a_{1}\sqrt{5} \, | \, a_{0}, a_{1} \in \mathbbm{Q}  \}. $$

\begin{de} 
A complex number is an \emph{algebraic integer} if it is a root of a monic (leading coefficient equal to 1) polynomial with coefficients in $\mathbbm{Z}$. \\
\end{de}

\begin{de}
The \emph{ring of integers of a number field} $K$ is the subset of its algebraic integers. It is denoted $O_{K}$. \\
\end{de}

\begin{prop}[\textit{e.g.} Marcus, \cite{Mar77} p.~15]
Let $m \in \mathbbm{Z}$ satisfy $m \equiv 1 \pmod{4}$ and let $K$ be the quadratic number field $\mathbbm{Q}(\sqrt{m})$. Then,
$$O_{K} = \left\{ \frac{a+b\sqrt{m}}{2} \, | \, a,b \in \mathbbm{Z} \right\}. $$
\end{prop}

\noindent
Applying this characterization to the case $K=\mathbbm{Q}(\sqrt{5})$ yields:
\begin{align*}
O_{\mathbbm{Q}(\sqrt{5})} &= \left\{ \frac{a+b\sqrt{m}}{2} \, | \, a,b \in \mathbbm{Z}, \quad  a \equiv b \pmod{2} \right\} \\
O_{\mathbbm{Q}(\sqrt{5})} &= \mathbbm{Z}[\phi].
\end{align*}
where $\phi$ is the golden ratio $\frac{1+\sqrt{5}}{2}$. 

\subsection{Arithmetic subgroups $\Gamma^{\{4,3,3,5\}}(I)$}

Since $\phi$ and its square root are algebraic numbers, $\mathbbm{Q}(\sqrt{\phi})$ is a number field. Its ring of integers is $\mathbbm{Z}[\sqrt{\phi}]$, and therefore every matrix of $\Gamma^{\{4,3,3,5\}}$ has its coefficients in the ring $\mathbbm{Z}[\sqrt{\phi}]$. \\

\begin{de}
A number field is \emph{totally real} if all its embeddings in $\mathbbm{C}$ are embeddings in $\mathbbm{R}$. \\
\end{de}

\noindent
In order to prove the same asymptotic behaviour of the code parameters $n$, $k$ and $d$ as in Refs~\cite{GL14} and \cite{Mur16}, we need to work with a totally real number field. Note that the totally real number field condition is not explicit in \cite{GL14} but it is implicitly used to show that their arithmetic group $\Gamma$ is discrete. However $\mathbbm{Q}(\sqrt{\phi})$ is not a totally real number field. Indeed $\sqrt{\phi}$ has minimal polynomial $X^{4} - X^{2} + 1$ which factorises as $(X - \sqrt{\phi})(X + \sqrt{\phi})\left(X - i\sqrt{\frac{\sqrt{5}-1}{2}}\right)\left(X + i\sqrt{\frac{\sqrt{5}-1}{2}}\right)$ and thus $\mathbbm{Q}(\sqrt{\phi})$ admits the two non-real embeddings determined by $\sqrt{\phi} \mapsto i\sqrt{\frac{\sqrt{5}-1}{2}}$ and by $\sqrt{\phi} \mapsto -i\sqrt{\frac{\sqrt{5}-1}{2}}$. We therefore conjugate matrices of $\Gamma^{\{4,3,3,5\}}$ in such a way that all their entries now belong to a totally real number field. Since matrix multiplication is defined through addition and multiplication of their entries, it is sufficient to ensure that the four matrices generating $\Gamma^{\{4,3,3,5\}}$ have their entries in a totally real number field. \\

Observe that $\left(\begin{smallmatrix}
1/\sqrt{\phi} & 0 \\
0 & 1
\end{smallmatrix}\right)\left(
\begin{smallmatrix}
\phi &  -\sqrt{\phi} \\
\sqrt{\phi} & -\phi 
\end{smallmatrix}\right)\left(
\begin{smallmatrix}
\sqrt{\phi} & 0 \\
0 & 1
\end{smallmatrix}\right) =
\left(\begin{smallmatrix}
\phi & -1 \\
\phi & -\phi
\end{smallmatrix}\right)$ and $\left(\begin{smallmatrix}
\phi & 1 \\
\phi & \phi
\end{smallmatrix}\right)^{-1} =
\left(\begin{smallmatrix}
\phi & -1 \\
\phi & -\phi
\end{smallmatrix}\right)$.
Therefore, defining $P = \mathrm{diag}(\sqrt{\phi}, 1, 1, 1, 1)$, the group $\tilde{\Gamma}^{\{4,3,3,5\}}$ defined as $P^{-1}\Gamma^{\{4,3,3,5\}} P$ has all its matrices with entries in the number field $K = \mathbbm{Q}(\phi) = \mathbbm{Q}(\sqrt{5})$, and even in its ring of integers $\mathbbm{Z}[\phi]$. Note that since $\Gamma^{\{4,3,3,5\}}$ is a subgroup of $O(1,4)$, matrices $g$ in $\tilde{\Gamma}^{\{4,3,3,5\}}$ satisfy the equation $g^{T}\tilde J g = \tilde J$ where $\tilde J = \text{diag}(-\phi,1,1,1,1)$. \\

Now the minimal polynomial of $\sqrt{5}$ is $X^{2} - 5$ which factorises as $(X - \sqrt{5})(X + \sqrt{5})$. Hence the two embeddings of $\mathbbm{Q}(\sqrt{5})$ in $\mathbbm{C}$ are the identity and the embedding determined by $\sqrt{5} \mapsto -\sqrt{5}$. $\mathbbm{Q}(\sqrt{5})$ is thus a totally real number field. \\

\begin{de}
Let $I$ be an ideal of a ring $A$. Let $G$ be a matrix group with coefficients in $A$. The \emph{principal congruence subgroup of level $I$ of G} is the kernel of the reduction modulo $I$ morphism. It is denoted $G(I) = \ker \pi_{I}$ with
\begin{align*}\pi_{I} : M_{n}(A) &\rightarrow M_{n}(A/I) \\
(a_{i,j}) &\mapsto (a_{i,j} + I).
\end{align*}
\end{de}

\noindent
It is natural to consider ideals of the ring $A$ because we want the quotient $A/I$ to be a ring in order for $M_{n}(A/I)$ to be defined. Hence to each ideal $I$ of $\mathbbm{Z}[\phi]$ corresponds a normal subgroup $\Gamma^{\{4,3,3,5\}}(I)$ of $\Gamma^{\{4,3,3,5\}}$. We denote by $\mathcal{M}(I)$ the quotient of $\mathbbm{H}^{4}$ by $\Gamma^{\{4,3,3,5\}}(I)$. By definition $\mathcal{M}(I) = \Gamma^{\{4,3,3,5\}}(I) \backslash \mathbbm{H}^{4}$ is the set of orbits of $\mathbbm{H}^{4}$ under the action of $\Gamma^{\{4,3,3,5\}}(I)$. Note that we use the notation $\Gamma^{\{4,3,3,5\}}(I) \backslash \mathbbm{H}^{4}$ and not  $\mathbbm{H}^{4}/\Gamma^{\{4,3,3,5\}}(I)$ because $\Gamma^{\{4,3,3,5\}}(I)$ acts on $\mathbbm{H}^{4}$ on the left. $\mathcal{M}(I)$ naturally inherits the hyperbolic structure of $\mathbbm{H}^{4}$. \\

For completeness, we will now detail how $\mathcal{M}(I)$ inherits the $\{4,3,3,5\}$ tessellation of $\mathbbm{H}^4$. By definition of $\Gamma(I)$ the following short sequence is exact:
$$ 1 \rightarrow \Gamma^{\{4,3,3,5\}}(I) \rightarrow \Gamma^{\{4,3,3,5\}} \rightarrow \pi_I(\Gamma^{\{4,3,3,5\}}) \rightarrow 1.$$

\noindent
Therefore, by the first isomorphism theorem, the quotient group $\Gamma^{\{4,3,3,5\}} / \Gamma^{\{4,3,3,5\}}(I)$ is isomorphic to $\pi_I(\Gamma^{\{4,3,3,5\}})$. This quotient group acts on $\mathcal{M}(I)$ in the following manner: for any $g \cdot \Gamma^{\{4,3,3,5\}}(I) \in \Gamma^{\{4,3,3,5\}} / \Gamma^{\{4,3,3,5\}}(I)$ and $\Gamma^{\{4,3,3,5\}}(I) \cdot x$ in $\mathcal{M}(I)$,
$$(g \cdot \Gamma^{\{4,3,3,5\}}(I)) \cdot (\Gamma^{\{4,3,3,5\}}(I) \cdot x) =  \Gamma^{\{4,3,3,5\}}(I) \cdot (g \cdot x).$$

\noindent
Since $\Gamma^{\{4,3,3,5\}}(I)$ is normal in $\Gamma^{\{4,3,3,5\}}$, this is well defined and it is a group action. We will see in subsection \ref{subsec:minimum_d} that for ideals with sufficiently large norms (see Definition \ref{norm}), $\Gamma^{\{4,3,3,5\}}(I)$ acts freely (without fixed points) on $\mathbbm{H}^4$. Therefore for such ideals $\mathcal{M}(I)$ is a manifold. \\

Moreover for $i \in \{0,\ldots,4\}$, $i$-faces of the $\{4,3,3,5\}$ tessellation of $\mathbbm{H}^4$ have a diameter upper bounded by some constant $c$ depending on the local structure. Again, for ideals with sufficiently large norms, $\Gamma^{\{4,3,3,5\}}(I)$ acts on $\mathbbm{H}^4$ in a way such that no pair of points $x, y \in \mathbbm{H}^4$ satisfying $d(x,y) \leq c$ belong to the same orbit. For such ideals $I$, the $\{4,3,3,5\}$ local structure is preserved by $\Gamma^{\{4,3,3,5\}}(I)$. We can retrieve it by considering cosets of $\pi_I(\Gamma^{\{4,3,3,5\}})$. \\

More precisely the group $\tilde{\Gamma}^{\{4,3,3,5\}}(I)$ is generated by $r_{0,\tilde{J}}, r_{1,\tilde{J}}, r_{2,\tilde{J}}, r_{3,\tilde{J}}$ and $r_{4,\tilde{J}}$:

\vspace{-10pt}
\begin{align*}
&r_{0,\tilde{J}} = \begin{pmatrix}
1 & 0 & 0 & 0 & 0\\
0 & -1 & 0 & 0 & 0\\
0 & 0 & 1 & 0 & 0\\
0 & 0 & 0 & 1 & 0\\
0 & 0 & 0 & 0 & 1\end{pmatrix}, \qquad
r_{1,\tilde{J}} = \begin{pmatrix}
1 & 0 & 0 & 0 & 0\\
0 & 0 & 1 & 0 & 0\\
0 & 1 & 0 & 0 & 0\\
0 & 0 & 0 & 1 & 0\\
0 & 0 & 0 & 0 & 1\end{pmatrix}, \qquad
r_{2,\tilde{J}} = \begin{pmatrix}
1 & 0 & 0 & 0 & 0\\
0 & 1 & 0 & 0 & 0\\
0 & 0 & 0 & 1 & 0\\
0 & 0 & 1 & 0 & 0\\
0 & 0 & 0 & 0 & 1\end{pmatrix}, \\
&r_{3,\tilde{J}} = \begin{pmatrix}
1 & 0 & 0 & 0 & 0\\
0 & 1 & 0 & 0 & 0\\
0 & 0 & 1 & 0 & 0\\
0 & 0 & 0 & 0 & 1\\
0 & 0 & 0 & 1 & 0\end{pmatrix}, \qquad
r_{4,\tilde{J}} = \begin{pmatrix}
\phi & 0 & 0 & 0 & -1 \\
0 & 1 & 0 & 0 & 0\\
0 & 0 & 1 & 0 & 0\\
0 & 0 & 0 & 1 & 0\\
\phi & 0 & 0 & 0 & -\phi \end{pmatrix}.
\end{align*}

\noindent
Therefore for any ideal $I$ of $\mathbbm{Z}[\phi]$, the group $\pi_I(\tilde{\Gamma}^{\{4,3,3,5\}})$ is generated by $r_{0,I}, r_{1,I}, r_{2,I}, r_{3,I}$ and $r_{4,I}$:

\begin{align*}
&r_{0,I} = \begin{pmatrix}
1 + I & 0 & 0 & 0 & 0\\
0 & -1 + I & 0 & 0 & 0\\
0 & 0 & 1 + I & 0 & 0\\
0 & 0 & 0 & 1 + I & 0\\
0 & 0 & 0 & 0 & 1 + I\end{pmatrix}, \qquad
&r_{1,I} = \begin{pmatrix}
1 + I & 0 & 0 & 0 & 0\\
0 & 0 & 1 + I & 0 & 0\\
0 & 1 + I & 0 & 0 & 0\\
0 & 0 & 0 & 1 + I & 0\\
0 & 0 & 0 & 0 & 1 + I\end{pmatrix}, \\
&r_{2,I} = \begin{pmatrix}
1 + I & 0 & 0 & 0 & 0\\
0 & 1 + I & 0 & 0 & 0\\
0 & 0 & 0 & 1 + I & 0\\
0 & 0 & 1 + I & 0 & 0\\
0 & 0 & 0 & 0 & 1 + I \end{pmatrix}, \qquad
&r_{3,I} = \begin{pmatrix}
1 + I & 0 & 0 & 0 & 0\\
0 & 1 + I & 0 & 0 & 0\\
0 & 0 & 1 + I & 0 & 0\\
0 & 0 & 0 & 0 & 1 + I\\
0 & 0 & 0 & 1 + I & 0\end{pmatrix}, \\
&r_{4,I} = \begin{pmatrix}
\phi + I & 0 & 0 & 0 & -1 + I \\
0 & 1 + I & 0 & 0 & 0\\
0 & 0 & 1 + I & 0 & 0\\
0 & 0 & 0 & 1 + I & 0\\
\phi + I & 0 & 0 & 0 & -\phi + I \end{pmatrix}. \\
\end{align*}

\noindent
For ideals $I$ whose norm is large enough, we can define $i$-faces of $\mathcal{M}(I)$ with the same coset method we used for $\mathbbm{H}^4$: $i$-faces correspond to cosets of $\pi_I(\tilde{\Gamma}^{\{4,3,3,5\}})$ by its subgroup $S_{i,I}$ generated by $(r_{j,I})_{j \neq i}$. Incident faces correspond to cosets whose intersection is not empty. We will use this method in Subsection \ref{subsection:small_ideals} to construct explicit quantum codes. \\

The results stated in the sequel of this paper are valid for ideals $I$ whose norm is large enough to have a $\{4,3,3,5\}$ tessellation of $\mathcal{M}(I)$. \\

\begin{de}
Let $H$ be a subgroup of a group $G$. The \emph{index of $H$ in $G$}, denoted $[G:H]$, is the cardinal of the quotient $G/H$. \\
\end{de}

\begin{lem}
The number of 2-faces of the $\{4,3,3,5\}$ tessellation of $\mathcal{M}(I)$ is \\
$n(I) = [\Gamma^{\{4,3,3,5\}}:\Gamma^{\{4,3,3,5\}}(I)] / 80$. \\
\end{lem}

\noindent
{\bf Proof:} $\mathcal{M}(I)$ admits a tessellation by $[\Gamma^{\{4,3,3,5\}}:\Gamma^{\{4,3,3,5\}}(I)]$ simplices isometric to a fundamental domain of the action of $\Gamma^{\{4,3,3,5\}}$ on $\mathbbm{H}^4$. A 2-face of the $\{4,3,3,5\}$ tessellation of $\mathcal{M}(I)$ corresponds to a coset of $S_{2,I}$ in $\Gamma^{\{4,3,3,5\}}(I)$. The result follows from the value of the cardinal of $S_{2,I}$: \\

\centerline{$|S_{2,I}| = |S_2| = |\langle r_0,r_1\rangle | \times |\langle r_3,r_4\rangle | = 8 \times 10 = 80.$}
 \hfill \square

\begin{de} \label{norm}
The \emph{norm} $N(I)$ of an ideal $I$ of a ring $A$ is the cardinal of the quotient $A/I$. \\
\end{de}

It is shown in Ref.~\cite{Mur16} that $[\Gamma^{\{4,3,3,5\}}:\Gamma^{\{4,3,3,5\}}(I)] \leq 4N(I)^{\text{dim}(O(1,4))} = 4N(I)^{10}$. This provides an upper bound on the size of the quantum code associated with an ideal $I$:

\vspace{-10pt}
\begin{equation}
\label{nN}
n(I) \leq N(I)^{10}/20.
\end{equation}

Note that the ring $\mathbbm{Z}[\phi]$ admits a family of ideals whose norms are unbounded. Indeed the norm of the ideal of $\mathbbm{Z}[\phi]$ generated by $m$ is $m^{2}$. This translates into a family of quantum codes with an unbounded number of physical qubits. Moreover there are other ideals in $\mathbbm{Z}[\phi]$. For example, the ideal generated by $\sqrt{5}$ has norm 5. \\

We will now paraphrase the correspondence exposed in Ref.~\cite{GL14} between a family of coverings and a family of quantum codes. From each 4-dimensional manifold equipped with a $\{4,3,3,5\}$ tessellation $\mathcal{M}(I)$, a code is constructed: qubits are identified with 2-faces of $\mathcal{M}(I)$, $X$-type stabilizers are identified with 1-faces (edges) of $\mathcal{M}(I)$ and $Z$-type stabilizers are identified with 3-faces of $\mathcal{M}(I)$. Each $X$-type, respectively $Z$-type, stabilizer acts by an $X$ Pauli matrix, respectively a $Z$ Pauli matrix, on every qubit it is incident to. The codespace is the common (+1)-eigenspace of the set of stabilizers. The length $n$ of the code, \textit{i.e.} its number of physical qubits, is the number of 2-faces of the tessellation. It is proportional to the volume of $\mathcal{M}(I)$. The dimension $k$ of the code, \textit{i.e.} its number of logical qubits, is the second Betti number of $\mathcal{M}(I)$, \textit{i.e.} the rank of its second homology group. The minimum distance $d$ of the code is the minimal number of 2-faces forming a homologically nontrivial 2-cycle in $\mathcal{M}(I)$. It is lower bounded by a quantity proportional to the least area of a homologically nontrivial surface of $\mathcal{M}(I)$. These proportionality coefficients do not depend on the ideal $I$. With this correspondence, the asymptotic behaviour of $n$, $k$ and $d$ is understood in terms of the family of manifolds $(\mathcal{M}(I))_{I \in \mathbbm{Z}[\phi]}$ independently of the $\{4,3,3,5\}$ tessellation.\\

To each ideal $I$ of the ring of integers $\mathbbm{Z}[\phi]$  corresponds a manifold $\mathcal{M}(I)$ equipped with a $\{4,3,3,5\}$ tessellation and a quantum error correcting code $\mathcal{C}(I)$.

\subsection{Lower bound on the rate of the quantum codes}

Quantum codes based on regular tessellations of hyperbolic spaces have a non-vanishing rate. It is well-known for tessellations of the hyperbolic plane and it is also true for tessellations of the hyperbolic 4-space. The argument is given in Ref.~\cite{GL14} (Theorem 7 and Corollary 9) and we can sketch it here and make it more quantitative than in Ref.~\cite{GL14}: \\

As a consequence of Gauss-Bonnet-Chern's theorem \cite{Che96}, the Euler characteristic $\chi(I)$ of the closed oriented hyperbolic 4-manifolds $\mathcal{M}(I)$ satisfies $\chi(I) = c \, \text{vol}(\mathcal{M}(I))$.
It is possible to generalise the definition of the Euler characteristic (see \textit{e.g.} \cite{Mar15})  to orbifolds (roughly speaking, manifolds that can have singularities) in such a way that this definition still holds. We can illustrate this by computing the Euler characteristic of the hypercube T of the \{4,3,3,5\} tessellation of hyperbolic 4-space. For each i in $\{0, \ldots,4\}$ we have to divide the number of i-faces of T by the number of hypercubes an i-face would be incident to in the \{4,3,3,5\} tessellation of hyperbolic 4-space. We obtain:

\vspace{-10pt}
\begin{align*}
\chi(T) &= \frac{1}{1} - \frac{8}{2} + \frac{24}{5} - \frac{32}{20} + \frac{16}{600} \\
&= \frac{17}{75}.
\end{align*}

\noindent
Gauss-Bonnet-Chern theorem also yields $\chi(T) = c \, \text{vol}(T)$.

\vspace{-10pt}
\begin{align*}
n(I) &= \#(\text{2-faces}) \\
&= \frac{24}{5}\#(\text{4-faces}) \\
&= \frac{24}{5} \frac{\text{vol}(\mathcal{M}(I))}{\text{vol}(T)} \\
&= \frac{24}{5} \frac{\chi(I)}{\chi(T)} \\
&= \frac{360}{17} \chi(I).
\end{align*}

\noindent
Moreover, by definition of the Euler characteristic, 
$\chi(I) = \sum_{i =0}^4 (-1)^i \dim H_{i}(\mathcal{M}(I),\mathbbm{Z}_{2})$, where $H_{i}(\mathcal{M}(I),\mathbbm{Z}_{2})$ is the $i^{\mathrm{th}}$ homology group of $\mathcal{M}(I)$ with coefficients in $\mathbbm{Z}_{2}$. \\
Since $\mathcal{M}(I)$ is a connected 4-manifold,  $\dim H_{0}(\mathcal{M}(I),\mathbbm{Z}_{2}) = \dim H_{4}(\mathcal{M}(I),\mathbbm{Z}_{2}) = 1$. \\
Since physical qubits are identified with 2-faces of the tessellation, the number of logical qubits $k(I)$ of the quantum code corresponding to $\mathcal{M}(I)$ is $\dim H_{2}(\mathcal{M}(I),\mathbbm{Z}_{2})$.

\vspace{-10pt}
\begin{align*}
k(I) &= \chi(I) + \dim H_{1}(\mathcal{M}(I),\mathbbm{Z}_{2}) + \dim H_{3}(\mathcal{M}(I)\mathbbm{Z}_{2}) - 2 \\
&\geq \chi(I) - 2 \\
&\geq \frac{17}{360} n(I) -2
\end{align*}

\noindent
This proves that the asymptotic rate of this family of quantum codes is greater than or equal to $\frac{17}{360} \approx 0.0472$. \\

Note that with the $\{5,3,3,5\}$ tessellation, the lower bound on the asymptotic rate is $\frac{5}{720} \times 26 \approx 0.18$. The other regular tessellations of hyperbolic 4-space yield lower rates.

\subsection{Lower bound on the minimum distances of the quantum codes}
\label{subsec:minimum_d} 

Following Ref.~\cite{GL14} we could prove that the minimum distance d asymptotically satisfies $n^{\epsilon} \leq d \leq n^{0.3}$ for an $\epsilon > 0$. But we will rather follow Ref.~\cite{Mur16} and Ref.~\cite{Mur17} and derive a tighter lower bound for the minimum distance: $d = \Omega(n^{0.1})$. We will also mention a variant of the construction yielding the even better $d = \Omega(n^{0.2})$. These two lower bounds rely on algebraic arguments.\\

The first lower bound on the minimum distance is obtained by lower-bounding the trace of matrices of $\tilde \Gamma^{\{4,3,3,5\}}(I)$. Indeed this lower bound on the trace of a matrix $g$ then yields a lower bound on the distance between a point $x \in \mathbbm{H}^{4}$ and its image $g \cdot x$. Finally, through Anderson's theorem (Ref.~\cite{GL14} th. 17) the size of the smallest homologically nontrivial 2-cycle is exponentially controlled by the size of the smallest homologically nontrivial 1-cycle. \\

We will start by deriving the lower bound on the trace of a matrix $g$ of $\tilde \Gamma^{\{4,3,3,5\}}(I)$. $g$ satisfies the matrix equation $g^{T} \tilde J g = \tilde J$ where $\tilde J = \text{diag}(-\phi,1,1,1,1)$. This translates into 10 quadratic equations on the entries of $g$. We will only need the five equations coming from entries on the diagonal:

\vspace{-10pt}
\begin{align*}
-\phi \, g_{0,0}^2 + g_{1,0}^2 + g_{2,0}^2 + g_{3,0}^2 + g_{4,0}^2 &= - \phi, \\
-\phi \, g_{0,j}^2 + g_{1,j}^2 + g_{2,j}^2 + g_{3,j}^2 + g_{4,j}^2 &= 1 \quad \text{for} \, j \in \{1,\ldots,4\}.
\end{align*}

\noindent
Denoting by $\sigma$ the nontrivial embedding of $\mathbbm{Q}(\sqrt{5})$ in $\mathbbm{C}$ that sends $\sqrt{5}$ to $-\sqrt{5}$ and applying it to these equations yields:

\begin{align}
-\sigma(\phi) \, \sigma(g_{0,0})^2 + \sigma(g_{1,0})^2 + \sigma(g_{2,0})^2 + \sigma(g_{3,0})^2 + \sigma(g_{4,0})^2 &= - \sigma(\phi), \label{diag0} \\
-\sigma(\phi) \, \sigma(g_{0,j})^2 + \sigma(g_{1,j})^2 + \sigma(g_{2,j})^2 + \sigma(g_{3,j})^2 + \sigma(g_{4,j})^2 &= 1 \quad \text{for} \, j \in \{1,\ldots,4\}. \label{diagj}
\end{align}

\noindent
Observing that $-\sigma(\phi)$ is positive, we obtain from Eq.(\ref{diag0}) that $|\sigma(g_{0,0})| \leq 1$. Similarly Eq.(\ref{diagj}) yields $|\sigma(g_{j,j})| \leq 1$ for $j \in \{1,\ldots,4\}$. Defining for $j \in \{0,\ldots ,4\}$, $y_{j} := g_{j,j} -1$, we have $|\sigma(y_{j})| \leq 2$ for $j \in \{0,\ldots,4\}$.
Moreover we can rewrite Eq.(\ref{diag0}) and Eq.(\ref{diagj}):

\begin{align}
-2\phi \, y_{0} -\phi \, y_{0}^2 + g_{1,0}^2 + g_{2,0}^2 + g_{3,0}^2 + g_{4,0}^2 &= 0, \label{y0} \\
2 y_{j} + y_{j}^{2} -\phi \, g_{0,j}^2 + \sum_{i \neq j} g_{i,j}^2  &= 0 \quad \text{for} \, j \in \{1,\ldots,4\}. \label{yj}
\end{align}

\noindent
From Eq.(\ref{y0}) and Eq.(\ref{yj}) we obtain that $2\phi \, y_{0}$ and $2 y_{j}$ for $j \in \{1,\ldots,4\}$ belong to $I^{2}$. \\

\begin{de}
The norm $N(x)$ of an element $x$ of a number field is the product of its conjugates. For a quadratic field with non trivial embedding $\sigma$, $N(x) = x \sigma(x)$. \\
\end{de} 

\begin{prop}[\textit{e.g.} \cite{Mur17} ]
The absolute value of the norm of an element of an ideal is greater than or equal to the norm of this ideal. \\
\end{prop}

\noindent
By multiplication and summation we know that $2 \phi (y_{0} + y_{1} + ... + y_{4})$ belongs to $I^{2}$. Hence,

\vspace{-10pt}
\begin{align*}
|N(2 \phi (y_{0} + y_{1} + ... + y_{4}))| &\geq N(I)^{2} \\
|N(y_{0} + y_{1} + ... + y_{4})| &\geq \frac{N(I)^{2}}{N(2 \phi)} \\
&\geq \frac{N(I)^{2}}{4}.
\end{align*}

\noindent
Therefore,

\vspace{-10pt}
\begin{align*}
|y_{0} + y_{1} + ... + y_{4}| &= \frac{|N(y_{0} + y_{1} + ... + y_{4})|}{|\sigma(y_{0} + y_{1} + ... + y_{4})|} \\
 &\geq \frac{\frac{N(I)^{2}}{4}}{|\sigma(y_{0})| + |\sigma(y_{1})| + ... + |\sigma(y_{4})|} \\
 &\geq \frac{N(I)^{2}}{40}.
\end{align*}

\noindent
Since $\text{tr}(g) = y_{0} + y_{1} + ... + y_{4} + 5$, we obtain $|\text{tr}(g)| \geq \frac{N(I)^{2}}{40} - 5.$ \\

Injecting Eq. \ref{nN}, we can lower-bound the absolute value of the trace of a matrix $g \in \Gamma^{\{4,3,3,5\}}(I)$ by the number of physical qubits $n(I)$:

\vspace{-10pt}
\begin{equation}
\nonumber
|\text{tr}(g)| \geq \frac{1}{2 \times 20^{0.8}} n(I)^{0.2} -5.
\end{equation}

We will now define the displacement function of a matrix $g \in \Gamma^{\{4,3,3,5\}}(I)$ acting on $\mathbbm{H}^{4}$ and lower-bound it by $|\text{tr}(g)|$. \\

\begin{de}
The displacement function $\rho$ of a matrix $M$ acting on a space $X$ is the infimum over $x \in X$ of the distance between $x$ and $Mx$: 
$$\rho_{M} = \inf_{x \in X} d(x,Mx).$$
\end{de}

\smallskip

\noindent
In our case, since the quotient manifold $\mathcal{M}(I)$ is closed and compact, the 1-systole is nothing but the infimum over $g \in \Gamma^{\{4,3,3,5\}}(I)$ of the displacement function of $g$: 
$$\text{1-syst}(\mathcal{M}(I)) = \inf_{g \in \Gamma^{\{4,3,3,5\}}(I)} \rho_{g}.$$

\noindent
Observing that both the trace and the displacement function are invariant by conjugation it is easy to prove (see \cite{Mur17} Proposition 6.1.1 p.64):
$$|\text{tr}(g)| \leq 2 \cosh(\rho_{g}) + 3.$$

\noindent
Thus,
\vspace{-10pt}
\begin{align*}
\text{1-syst}(\mathcal{M}(I)) & \geq \rho_g\\
& \geq \ln(|\text{tr}(g)| - 4) \\
&\geq \ln\left( \frac{1}{2 \times 20^{0.8}} n(I)^{0.2} -9 \right) \\
&\geq \ln\left( \frac{ n(I)^{0.2} - 18 \times 20^{0.8}}{2 \times 20^{0.8}} \right).
\end{align*}

\medskip

\begin{de}
The injectivity radius of a hyperbolic manifold is the supremum of the radii r such that the restriction of the covering projection $\mathbbm{H}^4 \rightarrow \mathcal{M}$ to any ball of radius r is injective. \\
\end{de}

\noindent
The injectivity radius $R(I)$ of the closed compact manifold $\mathcal{M}(I)$ is half its 1-systole:

\vspace{-10pt}
\begin{align*}
R(I) &\geq \frac{\text{1-syst}(\mathcal{M}(I))}{2}, \\
&\geq \ln \left( \left( \frac{ n(I)^{0.2} - 18 \times 20^{0.8}}{2 \times 20^{0.8}} \right)^{0.5} \right).
\end{align*}

\noindent
A specific case of Anderson's theorem yields:
\begin{thm}[\cite{GL14} th. 17] \label{Anderson}
Let $M$ be a closed manifold with a hyperbolic metric. Let $Z$ be a homologically non-trivial 2-cycle with coefficients in $\mathbbm{Z}_{2}$. Let $R$ be the injectivity radius of $M$. Then the volume of $Z$ is greater than or equal to the volume of a disk of radius $R$ in the hyperbolic plane:
$$\text{vol}(Z) \geq 2\pi (\cosh(R)-1).$$ 
\end{thm}

\medskip

\noindent
Since every 2-face of the \{4,3,3,5\} tessellation has the same volume $v$, for every 2-chain $C$ of $\mathcal{M}(I)$ with its tessellation, $\text{vol}(C) =  \text{wt}(C) \times v$ where wt($C$) is the number of faces of the chain $C$. To have a fully explicit result, we will compute the value of $v$, which is also the area of a square in the regular \{4,5\} tessellation of the hyperbolic plane. We can compute its value thanks to the (2-dimensional) Gauss-Bonnet theorem:

\vspace{-10pt}
\begin{align*}
v &= -2 \pi \, \chi(\{4,5\}\text{-square}) \\
 &= -2 \pi \left(\frac{1}{1} - \frac{4}{2} + \frac{4}{5}\right) \\
 &= \frac{2}{5} \pi.
\end{align*}

\noindent
Applying Theorem \ref{Anderson} to $\mathcal{M}(I)$ yields:
$$d(I) \geq \frac{\pi}{v} ( \exp(R(I)) -2 ),$$
where $d(I)$ is the minimal distance of the quantum code corresponding to $\mathcal{M}(I)$.

\noindent
This gives the bound on the minimal distance:
$$d(I) \geq \frac{5}{2} \left( \frac{ n(I)^{0.2} - 18 \times 20^{0.8}}{2 \times 20^{0.8}} \right)^{0.5} -5 .$$

\noindent
Asymptotically, $d(I)$ is greater than or equal to $\frac{5}{2^{1.5} \times 20^{0.4}} \, n(I)^{0.1} \approx 0.53 \, n(I)^{0.1}$.

\vspace{20pt}

Similarly to Ref.~\cite{Mur16}, we can consider the spin group $\text{Spin}(1,4)$, which is a double covering of $SO^{o}(1,4)$. Defining principal congruence subgroups at the level of the spin group $\text{Spin}(1,4)$, Murillo shows that the minimum distance $d$ of the corresponding codes satisfies $d = \Omega(n^{0.2})$ \cite{Mur16}. We note that the arithmetic manifolds defined at the level of the spin group are not strictly speaking the same as the ones defined at the level of the indefinite orthogonal group. Indeed the arithmetic subgroups of $\Gamma^{\{4,3,3,5\}}$ by which the hyperbolic 4-space is quotiented are different. To derive the lower bound $n^{0.2}$ on the minimum distance, the whole construction has to be done at the level of the spin group. Doing so does not alter the rate of the family of codes nor its $\{4,3,3,5\}$ local structure. Therefore it does not modify the local decoders designed in Section \ref{sec:decoding}. However since using the spin group makes the exposition less intuitive and does not improve the main result qualitatively, we will not state it in the main theorem: \\

\begin{thm} \label{main}
There exists a family of homological quantum error correcting codes $[[n,k,d]]$ defined from hyperbolic 4-manifolds equipped with $\{4,3,3,5\}$ tessellations. This family has non-vanishing rate $\frac{k}{n}$ which is asymptotically lower bounded by $17/360$. The minimum distance $d$ of its codes grows at least like $n^{0.1}$.
\end{thm}

\subsection{Estimates of the number of physical qubits}
\label{subsection:small_ideals}

The family of codes used to state Theorem \ref{main} has the drawback of being sparse. We show now that the smallest value of $n$ corresponding to a proper ideal of $\mathbbm{Z}[\phi]$ is 234 000. However there are normal subgroups of $\Gamma^{\{4,3,3,5\}}$ which are not constructed from an ideal of $\mathbbm{Z}[\phi]$. Finding such normal subgroups with small index in $\Gamma^{\{4,3,3,5\}}$ would lead to quantum codes with a more reasonable, \textit{i.e.} small enough to be practical, number of physical qubits. Even though the control over the minimum distance is lost when considering non arithmetic normal subgroups, the rate of the family of codes and the local decoders are valid for any normal subgroup. Moreover it could be interesting to use the technique of Ref.~\cite{BVCKT17} to interpolate between arithmetic hyperbolic 4-dimensional codes and \textit{e.g.} Euclidean 4-dimensional codes. This can be done by refining the hyperbolic tessellation by a Euclidean tessellation of the hypercubes. \\

Since $SO^{o}(1,4)$ has dimension 10, the number of hypercubes in the manifold equipped with a $\{4,3,3,5\}$ tessellation $\mathcal{M}(I)$ is proportional to $N(I)^{10}$. Therefore the number of qubits of the quantum error correcting code is also proportional to $N(I)^{10}$. In $\mathbbm{Z}[\phi]$, the smallest proper ideals we have found have norm 4, 5, 9, 11. The ideal whose norm is 4 is $2\mathbbm{Z}[\phi]$. But we have to ignore this ideal because $r_{0,2\mathbbm{Z}[\phi]}$ is the identity. The ideal whose norm is 5 is $\sqrt{5} \mathbbm{Z}[\phi]$. It corresponds to a number of qubits of the order of $5^{10} \approx 10^{7}$. \\

More precisely, we can compute an upper bound on the cardinal of $\pi_I(\Gamma^{\{4,3,3,5\}})$ for the ideal $I =\sqrt{5} \mathbbm{Z}[\phi]$. Since $\mathbbm{Z}[\phi]/(\sqrt{5}\mathbbm{Z}[\phi])$ is the finite field $\mathbbm{F}_5$, $\pi_{\sqrt{5}\mathbbm{Z}[\phi]}(\Gamma^{\{4,3,3,5\}})$ is isomorphic to a subgroup of the orthogonal group with dimension 5 and entries in $\mathbbm{F}_5$. Using the result of \cite{Wil09}, Sec. 3.7.2 p. 72, we obtain $|\pi_{\sqrt{5}\mathbbm{Z}[\phi]}(\Gamma^{\{4,3,3,5\}})| \leq 18\, 720\, 000$. Moreover, since $\phi + \sqrt{5}\mathbbm{Z}[\phi]$ in $\mathbbm{Z}[\phi]/(\sqrt{5}\mathbbm{Z}[\phi])$ corresponds to 3 in $\mathbbm{F}_5$, we have the following generating set for $\pi_{\sqrt{5}\mathbbm{Z}[\phi]}(\Gamma^{\{4,3,3,5\}})$:

\vspace{-10pt}
\begin{align*}
&r_{0,\mathbbm{F}_{5}} = \begin{pmatrix}
1 & 0 & 0 & 0 & 0\\
0 & 4 & 0 & 0 & 0\\
0 & 0 & 1 & 0 & 0\\
0 & 0 & 0 & 1 & 0\\
0 & 0 & 0 & 0 & 1\end{pmatrix}, \qquad
r_{1,\mathbbm{F}_{5}} = \begin{pmatrix}
1 & 0 & 0 & 0 & 0\\
0 & 0 & 1 & 0 & 0\\
0 & 1 & 0 & 0 & 0\\
0 & 0 & 0 & 1 & 0\\
0 & 0 & 0 & 0 & 1\end{pmatrix}, \qquad
r_{2,\mathbbm{F}_{5}} = \begin{pmatrix}
1 & 0 & 0 & 0 & 0\\
0 & 1 & 0 & 0 & 0\\
0 & 0 & 0 & 1 & 0\\
0 & 0 & 1 & 0 & 0\\
0 & 0 & 0 & 0 & 1\end{pmatrix}, \\
&r_{3,\mathbbm{F}_{5}} = \begin{pmatrix}
1 & 0 & 0 & 0 & 0\\
0 & 1 & 0 & 0 & 0\\
0 & 0 & 1 & 0 & 0\\
0 & 0 & 0 & 0 & 1\\
0 & 0 & 0 & 1 & 0\end{pmatrix}, \qquad
r_{4,\mathbbm{F}_{5}} = \begin{pmatrix}
3 & 0 & 0 & 0 & 4 \\
0 & 1 & 0 & 0 & 0\\
0 & 0 & 1 & 0 & 0\\
0 & 0 & 0 & 1 & 0\\
3 & 0 & 0 & 0 & 2 \end{pmatrix}.
\end{align*}

\noindent
Using the software GAP for computational discrete algebra and this set of generators, we find that $|\pi_{\sqrt{5}\mathbbm{Z}[\phi]}(\Gamma^{\{4,3,3,5\}})| = 18 \,720 \, 000$ (which implies that $\pi_{\sqrt{5}\mathbbm{Z}[\phi]}(\Gamma^{\{4,3,3,5\}})$ is the whole orthogonal group with dimension 5 and entries in $\mathbbm{F}_5$). Since $|S_{2,(\sqrt{5}\mathbbm{Z}[\phi])}| = 80$, the number $n(\sqrt{5}\mathbbm{Z}[\phi])$ of physical qubits of the corresponding code is $18 \,720 \,000/80 = 234\, 000$. Using a computational discrete algebra software like GAP and the coset method, we can compute the parity check matrices of this code.

\section{Local Decoders} 
\label{sec:decoding}

In this section, we design efficient decoding algorithms for the family of codes constructed in the previous section. These decoders are tailored for the whole hyperbolic 4-space equipped with a $\{4,3,3,5\}$ tessellation. Of course we want to apply these decoders to codes with a finite number of physical qubits, \textit{i.e.} to the hyperbolic manifolds $\mathcal{M}(I)$  equipped with a $\{4,3,3,5\}$ tessellation.

In this work we consider arbitrary errors of weight logarithmic in the number of physical qubits. Indeed the injectivity radii of the arithmetic hyperbolic manifolds associated with the golden code family scale logarithmically with their volumes. In terms of decoding, this implies that decoding a number of errors logarithmic in the number of physical qubits is strictly equivalent in the arithmetic hyperbolic manifolds and in the whole hyperbolic 4-space equipped with a $\{4,3,3,5\}$ tessellation. In other words our decoder provably succeeds for any error pattern of weight logarithmic in the number of physical qubits. Second, the same decoder will succeed with high probability to correct random error patterns of weight linear in the number of physical qubits, for instance if the qubits are affected independently by depolarizing noise.

The advantage of our decoders over the generic hyperbolic 4-dimensional decoder by Hastings \cite{Has16} is their high locality. Indeed Hastings' decoder is local at the level of a ball of radius $R_{dec}$ where $R_{dec}$ is constant but unknown. Since in hyperbolic 4-space the number of 2-faces in a ball of radius $R_{dec}$ grows like $e^{3R_{dec}}$, even small values of $R_{dec}$ can lead to an unpractical degree of locality. For instance the authors of \cite{BDMT16} use the value $R_{dec} = 1.5$ to implement a version of Hastings' decoder in a 4-dimensional toric code setting. With such a small value of $R_{dec}$ the analysis of the performance of Hastings' decoder probably does not apply. 
The analysis of our decoders, on the other hand, is valid at a level of locality that is computationally practical. \\

Since the codes we consider are CSS, it is possible to decode $X$-type and $Z$-type errors independently, and this is what our algorithm does. Because correcting these two types of errors on a qubit is sufficient to correct an arbitrary single-qubit error, we can state our decoding theorem as follows. \\

\begin{thm} \label{dec}
There exists a constant $C$ such that for any error $E$ corrupting less than $C \log n$ physical qubits, the decoding algorithm returns a set of qubits $E'$ such that $E$ and $E'$ differ by a sum of stabilizers. \\
\end{thm}

Since stabilizers act trivially on the codespace, Theorem \ref{dec} implies that any codestate corrupted on at most $C \log n$ physical qubits is perfectly recovered by the active error correction procedure. \\

Moreover, standard results in percolation theory show that for a random error model where each qubit is affected independently and identically with a depolarizing node, then below some constant noise threshold, the error will affect qubits that belong to small connected components of the tessellation of size $O(\log n)$. This is because the tessellation has constant degree. Using the same ideas as in \cite{FGL18a}, the decoding algorithm will correct the error with high probability. \\

\begin{thm}
There exists a constant $p_0 >0$ such that if each qubit is independently and identically affected by an $X$ or a $Z$ error with probability $p < p_0$, then the decoding algorithm corrects the error with high probability.
\end{thm}

\subsection{Decoding $Z$-errors} \label{decoding Z}

As mentioned, the algorithm successively decodes $Z$-errors then $X$-errors. It succeeds if it recovers the right error patterns, up to some element of the stabilizer group.
We first consider $Z$-errors. A $Z$-decoder takes as input a syndrome on $X$-type stabilizers and outputs a set of $Z$-errors consistent with this syndrome. For golden codes, $X$-type stabilizers are defined by edges in the $\{4,3,3,5\}$ tessellation. The error pattern is by definition the set of 2-faces corresponding to qubits having a $Z$-error. The syndrome is the boundary of the error pattern. Since every boundary is a cycle, the syndrome consists of several loops of edges. \\

\begin{de}
A path of edges from vertex $v_{1}$ to vertex $v_{2}$ is \emph{minimal} if no other path of edges from vertex $v_{1}$ to vertex $v_{2}$ is shorter. \\
\end{de}

\noindent
The $Z$-decoder follows from following lemma:

\begin{lem} \label{Z loop}
In the $\{4,3,3,5\}$ tessellation, every loop of edges  has at least one subpath of length at most 8 which is not minimal.
\end{lem}
Lemma \ref{Z loop} is proven in the appendix. \\

\noindent
With Lemma \ref{Z loop} at hand, it is now easy to design a local decoder:

\begin{itemize}
\item From every edge of the syndrome, explore every path of edges in the syndrome of length at most 8.
\item If such a path is not minimal, flip qubits to decrease its length.
\item Iterate, until no non-minimal path of length at most 8 can be found.
\end{itemize}

While the complexity of the $Z$-decoder appears at first sight to be quadratic in the size of the syndrome, it can be made linear if one only explores in the $(i+1)^{\mathrm{th}}$ step paths that were not already explored during the $i$-th round of the algorithm. Indeed flipping a qubit only affects a constant number of paths of length at most 8. Moreover, as long as the error weight is below the injectivity radius of the manifold, or if the error consists of many such small connected components, then the syndrome weight is proportional to the error weight. This fact comes from the hyperbolicity of the tessellation.
In other words, the decoding algorithm has a complexity linear in the error weight. 

\subsection{Decoding $X$-errors} \label{decoding X}

We now turn our attention to decoding $X$-errors. An $X$-decoder takes as input a syndrome on $Z$-type stabilizers and outputs a set of $X$-errors consistent with this syndrome. For golden codes, $Z$-type stabilizers are defined by polyhedrons (3-faces) in the $\{4,3,3,5\}$ tessellation. It is more convenient for us to work with edges than with polyhedrons. We therefore consider the $\{5,3,3,4\}$ dual tessellation. With this point of view, $Z$-type stabilizers are defined by edges in the $\{5,3,3,4\}$ dual tessellation. \\

\noindent
The $X$-decoder follows from the following lemma:

\begin{lem} \label{X loop}
In the $\{5,3,3,4\}$ tessellation, every loop of edges has at least one subpath incident to a single 4-face and which is not minimal.
\end{lem}

\noindent
Lemma \ref{X loop} is proven in the appendix. \\

\noindent
With Lemma \ref{X loop} at hand, it is now easy to design a local decoder:

\begin{itemize}
\item From every edge of the syndrome, explore every path of edges in the syndrome incident to a single 4-face.
\item If such a path is not minimal, flip qubits to decrease its length.
\item Iterate, until no non-minimal path incident to a single 4-face can be found. 
\end{itemize}

\noindent
The complexity of this $X$-decoder is linear in the size of the error for the same reason as the $Z$-decoder.

\section{Conclusion and Perspectives}

In this work, we have presented a variant of the quantum LDPC code family due to Guth and Lubotzky. Like theirs, our family is also obtained by considering tessellations of hyperbolic 4-space, but the crucial new feature of our construction is that the tessellation is regular. We then exploit this regularity to design an efficient and explicit decoding algorithm that provably corrects arbitrary errors of weight $O(\log n)$ and decodes with high probability random independent and identically distributed errors provided the error rate is below some constant threshold. \\

We note that both the dimension 4 and hyperbolicity present advantages for decoding. Placing the qubits on 2-faces yields syndromes which are cycles of edges (or of coedges) and a decoder should simply try to shorten such cycles, which can be done efficiently by means of a local algorithm as we demonstrated. This algorithm is also more
efficient in hyperbolic space since the syndrome weight increases linearly with the error weight (for small errors). This is arguably simpler than pairing vertices as required in surface codes. Another advantage of 1-dimensional syndromes is that they contain redundant information, which should be helpful when considering more realistic scenarios where syndrome measurements are not assumed to be ideal. \\

Future work should focus on simulating the performance of hyperbolic 4-dimensional codes with respect to different error models. Although the code family based on quotienting by arithmetic subgroups is arguably out of reach for simulations, it will be interesting to consider quotienting by different normal subgroups, yielding codes of more reasonable size. While the bounds on the minimum distance would not apply anymore in that case, we expect the behaviour of the decoding algorithm to be essentially identical for the usual error models.

\section*{Acknowledgements}
We thank Gilles Z{\'e}mor for introducing the authors to the construction of \cite{GL14} and for fruitful discussions on tessellations and homological codes. We thank Nicolas Bergeron for mentioning the arithmeticity of the discrete isometry groups corresponding to regular tessellations of hyperbolic 4-space. We also thank Benjamin Audoux, Alain Couvreur, Antoine Grospellier,  Anirudh Krishna and Jean-Pierre Tillich for useful discussions on quantum codes.  
We are also truly grateful for the very detailed comments provided by an anonymous referee and which helped us to significantly improve our exposition.
 We acknowledge support from the ANR through the QuantERA project QCDA.

\appendix{: Proof of the $Z$-decoder Lemma}

\noindent
Before proving Lemma \ref{Z loop}, we first establish a 2-dimensional version of it. Even though this 2-dimensional version is irrelevant to decoding homological quantum codes, it allows us to illustrate the main ideas with figures and may help the reader understand the key role of hyperbolicity in Lemma \ref{Z loop}. \\

\begin{lem} \label{2D Z loop}
In the $\{4,5\}$ tessellation of hyperbolic plane, every loop of edges has at least one subpath of length at most 4 which is not minimal. \\
\end{lem} 

\begin{figure} [htbp]
\centerline{ \includegraphics[width=0.8\textwidth]{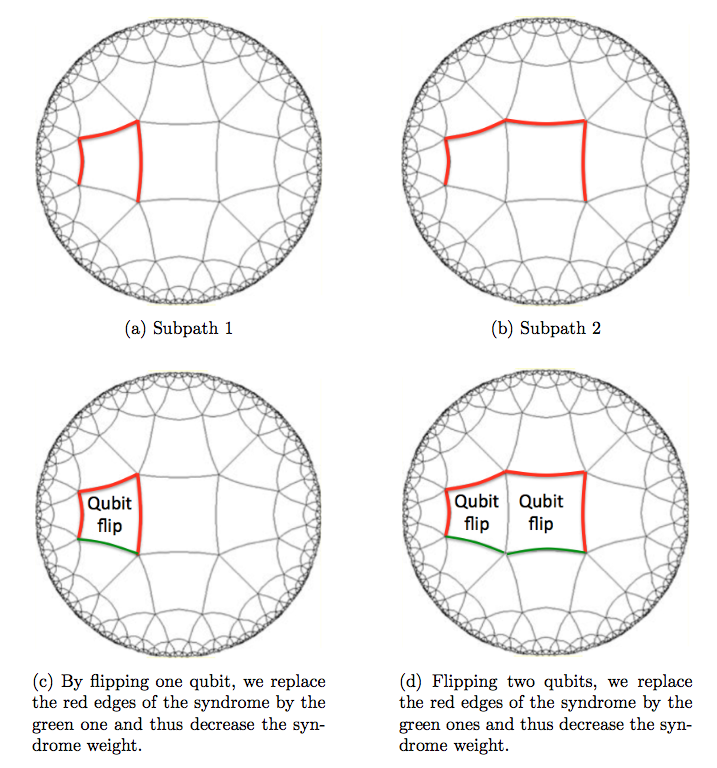}}
\vspace*{13pt}
\fcaption{In the $\{4,5\}$ tessellation of hyperbolic plane, every loop of edges contains one of the two subpaths in red. These two subpaths are not minimal: they can be replaced by the shorter green ones by flipping one or two qubits.  (source for image: \cite{tess})}
\label{4_5}
\end{figure}

\noindent
{\bf Proof:} Equivalently, in the $\{4,5\}$ tessellation of hyperbolic plane every loop of edges admits at least one of the two subpaths depicted on Fig. \ref{4_5}. 

\begin{figure} [htbp]
\centerline{ \includegraphics[width=0.8\textwidth]{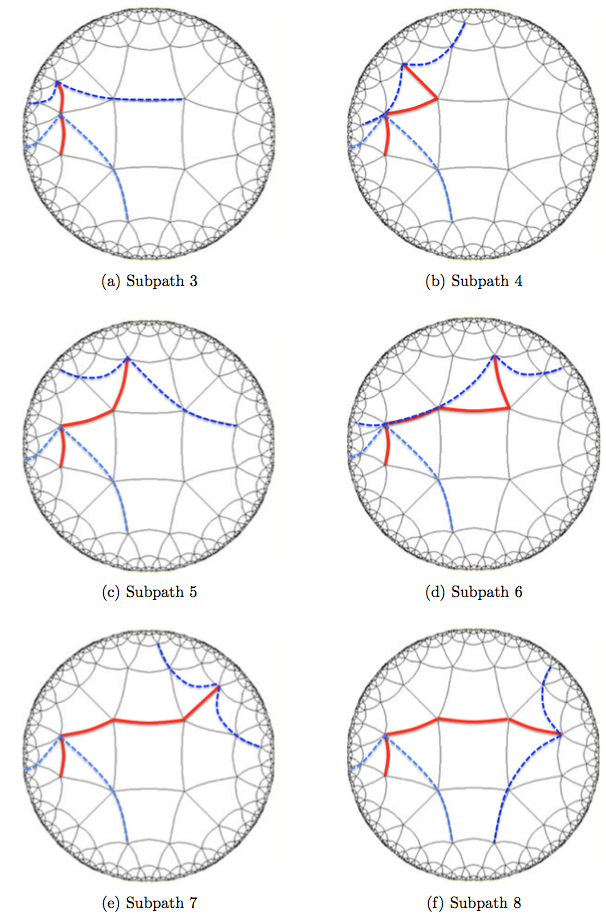}}
\vspace*{13pt}
\fcaption{The dark blue cone assigned to the last edge of the path contains the light blue cone assigned to the first edge of the path. Every minimal path of length 4 contains one of these six subpaths (or a subpath symmetric to it). Therefore if every length 4 subpath is minimal, it is impossible to form a loop. (source for image: \cite{tess})}
\label{cone}
\end{figure}

\noindent
It is sufficient to prove it on a single loop of edges of the $\{4,5\}$ tessellation of hyperbolic plane. We choose an arbitrary orientation on this loop.  An edge $e$ is written $e = \{v_{1},v_{2}\}$ if it is oriented from $v_{1}$ to $v_{2}$. To each edge $e = \{v_{1},v_{2}\}$ we assign a cone $C_{e}$ defined as the set of points of hyperbolic plane closer to $e$ than to any other edge incident to  $v_{2}$. The cone $C_{e}$ divides the hyperbolic plane in two regions: the outside of the cone and the inside of the cone. \\
We suppose by contradiction that there exists a loop $L$ of edges in the $\{4,5\}$ tessellation of hyperbolic plane such that every subpath of $L$ of length at most 4 is minimal. Figure \ref{cone} shows by an exhaustive search that for any edge $e$, there exists $f$ in $L \setminus \{e\}$ such that $C_{f}$ contains $C_{e}$. By immediate induction, it is then possible to construct a sequence $(e_{i})_{i \in \mathbbm{N}}$ of edges in $L$ such that $j > i$ implies that $C_{e_{j}}$ contains $C_{e_{i}}$. This contradicts the fact that $L$ is a loop. \hfill \square \\

\noindent
We can now prove Lemma \ref{Z loop}. \\
{\bf Proof }[of  Lemma \ref{Z loop}]:
It follows the exact same line. To each edge $e = \{v_{1},v_{2}\}$, we assign a cone $C_{e}$ defined as the set of points of hyperbolic 4-space closer to $e$ than to any other edge incident to  $v_{2}$. Since the number of length 8 paths on the edge graph of the $\{4,3,3,5\}$ tessellation is too high to check every case manually, we used a computer program to find that every minimal path of length 8 contains a subpath such that the cone assigned to its last edge contains the cone assigned to its first edge. Therefore in order to form a loop, at least one subpath of length at most 8 has to not be minimal. The decoder consists in flipping qubits in order to shorten this subpath.  \hfill \square

\newpage

\appendix{: Proof of the $X$-decoder Lemma}

\noindent
Before proving Lemma \ref{X loop} we prove a 2-dimensional version of it. Even though the 2-dimensional version is irrelevant to decoding homological quantum codes, it allows us to illustrate the main ideas with figures and may help the reader understand the key role of hyperbolicity in Lemma \ref{X loop}. \\

\begin{lem} \label{2D X loop}
In the $\{5,4\}$ tessellation of hyperbolic plane, every loop of edges admits at least one subpath incident to a single pentagon and which is not minimal. \\
\end{lem} 

\noindent
Equivalently, in the $\{5,4\}$ tessellation of hyperbolic plane every loop of edges has at least one subpath consisting of three edges incident to the same pentagon. After flipping the qubit corresponding to this pentagon, this subpath of length 3 (or 4) is replaced by a subpath of length 2 (or 1) and thus the syndrome weight is reduced. \\

\begin{figure} [htbp]
\centerline{ \includegraphics[width=0.8\textwidth]{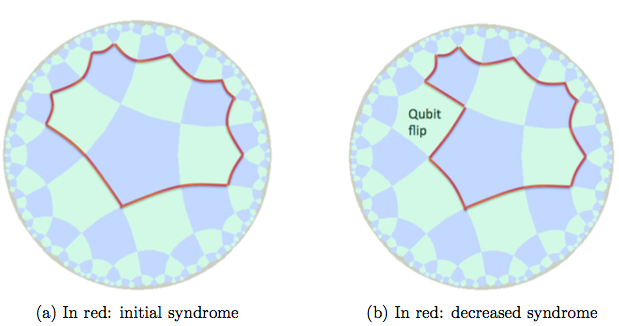}}
\vspace*{13pt}
\fcaption{Every loop of edges in the  $\{5,4\}$ tessellation of hyperbolic plane contains a subpath of three edges incident to the same pentagon. Flipping the qubit corresponding to this pentagon reduces the syndrome weight. (source for image: \cite{tess})}
\label{plane_syndrome}
\end{figure}

\begin{figure} [htbp]
\centerline{ \includegraphics[width=0.8\textwidth]{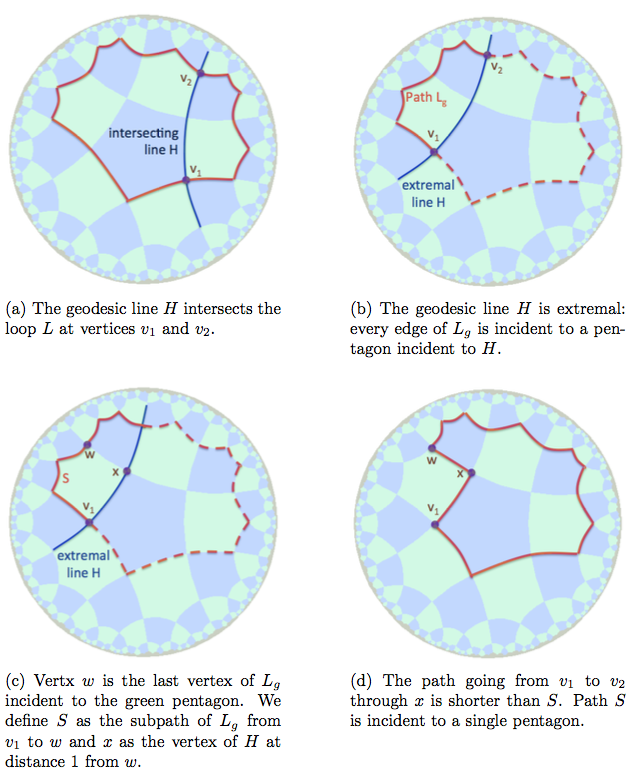}}
\vspace*{13pt}
\fcaption{ Every loop $L$ in the $\{5,4\}$ tessellation has a subpath $S$ consisting of three edges incident to the same pentagon. $S$ is not minimal since it can be replaced by a path of length 2. A similar property holds for loops in the $\{5,3,3,4\}$ tessellation. (source for image: \cite{tess}) }
\label{pentagon}
\end{figure}

\noindent
{\bf Proof:} We consider a loop $L$ of edges in the $\{5,4\}$ tessellation. \\
As shown in Fig. \ref{pentagon}(a), there exists a geodesic line $H$ in the $\{5,4\}$ tessellation which intersects the loop $L$ at two of its vertices $v_{1}$ and $v_{2}$. $v_{1}$ and $v_{2}$ define a partition of $L$ into two subpaths. We denote these two subpaths by $L_{g}$ and $L_{r}$. Without loss of generality, assume that the geodesic line $H$ is extremal with respect to $L_{g}$ in the sense that every edge in $L_{g}$ is incident to a pentagon incident to an edge of $H$. This is illustrated in Fig \ref{pentagon}(b). Without loss of generality, assume that the edge of $L_{g}$ incident to $v_{1}$ does not belong to the extremal geodesic line $H$. \\
If there exists a pentagon $P$ such that every edge in $L_{g}$ is incident to $P$, then $L_{g}$ is not minimal because the path in the geodesic line $H$ going from $v_{1}$ to $v_{2}$ consists of a single edge. It is thus shorter than $L_{g}$ and Lemma \ref{2D X loop} is proven in this case. \\
If such a pentagon $P$ doesn't exist, we denote by $w$ the last vertex of $L_{g}$ such that every vertex between $v_{1}$ and $w$ in $L_{g}$ is incident to a single pentagon (see Fig. \ref{pentagon}(c)). We consider the subpath $S$ of $L_{g}$ going from $v_{1}$ to $w$. $S$ is incident to a single pentagon. It has length 3. We denote by $x$ the vertex of $H$ at edge-distance 1 from $w$. The path $S'$ consisting of the edge $\{v_{1},x\}$ and the edge $\{x,w\}$ has length 2 (see Fig. \ref{pentagon}(d)). It is shorter than $S$. \hfill \square \\

\noindent
We are now ready to prove Lemma \ref{X loop}. \\
{\bf Proof }[of Lemma \ref{X loop}]:
The proof is very similar to the proof of Lemma \ref{2D X loop}. We consider a loop $L$ of edges in the $\{5,3,3,4\}$ tessellation. \\
As shown in Fig. \ref{pentagon}(a), there exists a geodesic hyperplane $H$ in the $\{5,3,3,4\}$ tessellation which intersects the loop $L$ at two of its vertices $v_{1}$ and $v_{2}$. Vertices $v_{1}$ and $v_{2}$ define a partition of $L$ into two subpaths. We denote these two subpaths by $L_{g}$ and $L_{r}$. Without loss of generality, assume that the geodesic hyperplane $H$ is extremal with respect to $L_{g}$ in the sense that every edge in $L_{g}$ is incident to a 4-face incident to an edge of $H$. This is illustrated in Fig. \ref{pentagon}(b). Without loss of generality, assume that the edge of $L_{g}$ incident to $v_{1}$ does not belong to the extremal geodesic hyperplane $H$. \\
If there exists a 4-face $P$ such that every edge in $L_{g}$ is incident to $P$, then $L_{g}$ is not minimal. Indeed the path in the geodesic hyperplane $H$ going from $v_{1}$ to $v_{2}$ is shorter than $L_{g}$ and Lemma \ref{X loop} is proven in this case. \\
If such a 4-face $P$ doesn't exist, we denote by $w$ the last vertex of $L_{g}$ such that every vertex between $v_{1}$ and $w$ in $L_{g}$ is incident to a single 4-face (see Fig. \ref{pentagon}(c)). We consider the subpath $S$ of $L_{g}$ going from $v_{1}$ to $w$. $S$ is incident to a single 4-face. We denote by $x$ the vertex of $H$ at edge-distance 1 from $w$. We define $S'$ as one of the shortest path in $H$ going from $v_{1}$ to $x$ concatenated with the single edge path  going from $x$ to $w$ (see Fig. \ref{pentagon}(d)). An exhaustive search on the 1-skeleton of a 120-cell shows that $S'$ is always shorter than $S$. \hfill \square

\vspace{50pt}

\nonumsection{References}
 
\renewenvironment{thebibliography}[1]
        {\frenchspacing
         \small\rm\baselineskip=11pt
         \begin{list}{\arabic{enumi}.}
        {\usecounter{enumi}\setlength{\parsep}{0pt}     
          \setlength{\leftmargin}{12.7pt}
                \setlength{\rightmargin}{0pt}
         \setlength{\itemsep}{0pt} \settowidth
          {\labelwidth}{#1.}\sloppy}}{\end{list}}


\begin{thebibliography}{000}

\bibitem{GL14}
Larry Guth and Alexander Lubotzky.
\newblock Quantum error correcting codes and 4-dimensional arithmetic
hyperbolic manifolds.
\newblock {\em Journal of Mathematical Physics}, 55(8):082202, 2014.

\bibitem{Sho95}
Peter~W Shor.
\newblock Scheme for reducing decoherence in quantum computer memory.
\newblock {\em Physical review A}, 52(4):R2493, 1995.

\bibitem{CS96}
A~Robert Calderbank and Peter~W Shor.
\newblock Good quantum error-correcting codes exist.
\newblock {\em Physical Review A}, 54(2):1098, 1996.

\bibitem{Got97}
Daniel Gottesman.
\newblock Stabilizer codes and quantum error correction.
\newblock {\em arXiv preprint quant-ph/9705052}, 1997.

\bibitem{KIT03}
A~Yu Kitaev.
\newblock Fault-tolerant quantum computation by anyons.
\newblock {\em Annals of Physics}, 303(1):2--30, 2003.

\bibitem{BM07}
Hector Bombin and Miguel~A Martin-Delgado.
\newblock Homological error correction: Classical and quantum codes.
\newblock {\em Journal of mathematical physics}, 48(5):052105, 2007.

\bibitem{Zem09}
Gilles Z{\'e}mor.
\newblock On cayley graphs, surface codes, and the limits of homological coding
for quantum error correction.
\newblock In {\em International Conference on Coding and Cryptology}, pages
259--273. Springer, 2009.

\bibitem{BT16}
Nikolas~P Breuckmann and Barbara~M Terhal.
\newblock Constructions and noise threshold of hyperbolic surface codes.
\newblock {\em IEEE Transactions on Information Theory}, 62(6):3731--3744,
2016.

\bibitem{DKLP02}
Eric Dennis, Alexei Kitaev, Andrew Landahl, and John Preskill.
\newblock Topological quantum memory.
\newblock {\em Journal of Mathematical Physics}, 43(9):4452--4505, 2002.

\bibitem{Har04}
James~William Harrington.
\newblock {\em Analysis of quantum error-correcting codes: symplectic lattice
codes and toric codes}.
\newblock PhD thesis, California Institute of Technology, 2004.

\bibitem{DZ17}
Nicolas Delfosse and Gilles Z{\'e}mor.
\newblock Linear-time maximum likelihood decoding of surface codes over the
quantum erasure channel.
\newblock {\em arXiv preprint arXiv:1703.01517}, 2017.

\bibitem{DN17}
Nicolas Delfosse and Naomi~H Nickerson.
\newblock Almost-linear time decoding algorithm for topological codes.
\newblock {\em arXiv preprint arXiv:1709.06218}, 2017.

\bibitem{FML02}
Michael~H Freedman, David~A Meyer, and Feng Luo.
\newblock Z2-systolic freedom and quantum codes.
\newblock {\em Mathematics of quantum computation, Chapman \& Hall/CRC}, pages
287--320, 2002.

\bibitem{BPT10}
Sergey Bravyi, David Poulin, and Barbara Terhal.
\newblock Tradeoffs for reliable quantum information storage in 2d systems.
\newblock {\em Physical review letters}, 104(5):050503, 2010.

\bibitem{Del13}
Nicolas Delfosse.
\newblock Tradeoffs for reliable quantum information storage in surface codes
and color codes.
\newblock In {\em Information Theory Proceedings (ISIT), 2013 IEEE
International Symposium on}, pages 917--921. IEEE, 2013.

\bibitem{Got13}
Daniel Gottesman.
\newblock Fault-tolerant quantum computation with constant overhead.
\newblock {\em arXiv preprint arXiv:1310.2984}, 2013.

\bibitem{FGL18}
Omar Fawzi, Antoine Grospellier, and Anthony Leverrier.
\newblock Constant overhead quantum fault-tolerance with quantum expander
codes.
\newblock In {\em 2018 IEEE 59th Annual Symposium on Foundations of Computer
Science (FOCS)}, pages 743--754. IEEE, 2018.

\bibitem{Has16}
MB~Hastings.
\newblock Quantum codes from high-dimensional manifolds.
\newblock {\em arXiv preprint arXiv:1608.05089}, 2016.

\bibitem{BDMT16}
Nikolas~P Breuckmann, Kasper Duivenvoorden, Dominik Michels, and Barbara~M
Terhal.
\newblock Local decoders for the 2d and 4d toric code.
\newblock {\em arXiv preprint arXiv:1609.00510}, 2016.

\bibitem{Rat06}
John Ratcliffe.
\newblock {\em Foundations of hyperbolic manifolds}, volume 149.
\newblock Springer Science \& Business Media, 2006.

\bibitem{MS02}
Peter McMullen and Egon Schulte.
\newblock {\em Abstract regular polytopes}, volume~92.
\newblock Cambridge University Press, 2002.

\bibitem{Cox54}
Harold Stephen~Macdonald Coxeter.
\newblock Regular honeycombs in hyperbolic space.
\newblock In {\em Proceedings of the International Congress of Mathematicians},
volume~3, pages 155--169, 1954.

\bibitem{Mar77}
Daniel~A Marcus.
\newblock {\em Number fields}, volume~8.
\newblock Springer, 1977.

\bibitem{Mur16}
Plinio~GP Murillo.
\newblock Systole of congruence coverings of arithmetic hyperbolic manifolds.
\newblock {\em arXiv preprint arXiv:1610.03870}, 2016.

\bibitem{Che96}
Shiing-shen Chern.
\newblock A simple intrinsic proof of the gauss-bonnet formula for closed
riemannin manifolds.
\newblock In {\em A Mathematician And His Mathematical Work: Selected Papers of
  SS Chern}, pages 115--120. World Scientific, 1996.

\bibitem{Mar15}
Bruno Martelli.
\newblock Hyperbolic four-manifolds.
\newblock {\em arXiv preprint arXiv:1512.03661}, 2015.

\bibitem{Mur17}
Plinio~GP Murillo.
\newblock {\em On Arithmetic Manifolds with Large Systole}.
\newblock PhD thesis, PhD thesis, IMPA, 2017.

\bibitem{BVCKT17}
Nikolas~P Breuckmann, Christophe Vuillot, Earl Campbell, Anirudh Krishna, and
  Barbara~M Terhal.
\newblock Hyperbolic and semi-hyperbolic surface codes for quantum storage.
\newblock {\em arXiv preprint arXiv:1703.00590}, 2017.

\bibitem{Wil09}
Robert Wilson.
\newblock {\em The finite simple groups}, volume 251.
\newblock Springer Science \& Business Media, 2009.

\bibitem{FGL18a}
Omar Fawzi, Antoine Grospellier, and Anthony Leverrier.
\newblock Efficient decoding of random errors for quantum expander codes.
\newblock In {\em Proceedings of the 50th Annual ACM SIGACT Symposium on Theory
  of Computing}, pages 521--534. ACM, 2018.

\bibitem{tess}
https://mathcs.clarku.edu/~djoyce/poincare/poincare.html
\newblock Accessed: 2010-09-30.

\end{thebibliography}
\end{document}